\documentclass[aps,pre,twocolumn,superscriptaddress,noshowpacs,longbibliography]{revtex4-2}
\usepackage{amsmath}
\usepackage{amssymb}
\usepackage{tikz}
\usepackage{graphicx}
\usepackage{dcolumn}
\usepackage{bm}
\usepackage{epsfig}
\usepackage{psfrag}
\def\II{\hbox{$1\hskip -1.2pt\vrule depth 0pt height 1.6ex width 0.7pt\vrule depth 0pt height 0.3pt width 0.12em$}}
\newcommand{\reffig}[1]{\mbox{Fig.~\ref{#1}}}
\newcommand{\refeq}[1]{\mbox{Eq.~(\ref{#1})}}
\newcommand{\refsec}[1]{\mbox{Sec.~\ref{#1}}}

\newcommand{\be}{\begin{equation}}
\newcommand{\ee}{\end{equation}}
\newcommand{\bal}{\begin{align}}
\newcommand{\eal}{\end{align}}
\newcommand{\ba}{\begin{eqnarray}}
\newcommand{\ea}{\end{eqnarray}}
\renewcommand{\Re}{\mathrm{Re}}
\renewcommand{\Im}{\mathrm{Im}}
\newcommand{\T}{${\mathcal T}\,$}
\newcommand{\Ti}{${\mathcal T}$}
\def\II{\hbox{$1\hskip -1.2pt\vrule depth 0pt height 1.6ex width 0.7pt\vrule depth 0pt height 0.3pt width 0.12em$}}

\begin{document}

\title{\bf Experimental test of the Rosenzweig-Porter model for the transition from Poisson to Gaussian unitary ensemble statistics}
\author{Xiaodong Zhang}
\address{%
Lanzhou Center for Theoretical Physics and the Gansu Provincial Key Laboratory of Theoretical Physics, Lanzhou University, Lanzhou, Gansu 730000, China
}
\author{Weihua Zhang}
\address{%
Lanzhou Center for Theoretical Physics and the Gansu Provincial Key Laboratory of Theoretical Physics, Lanzhou University, Lanzhou, Gansu 730000, China
}
\address{Center for Theoretical Physics of Complex Systems, Institute for Basic Science (IBS), Daejeon 34126, Korea}
\author{Jiongning Che}
\address{%
Lanzhou Center for Theoretical Physics and the Gansu Provincial Key Laboratory of Theoretical Physics, Lanzhou University, Lanzhou, Gansu 730000, China
}
\author{Barbara Dietz}
\email{bdietzp@gmail.com}
\address{%
Lanzhou Center for Theoretical Physics and the Gansu Provincial Key Laboratory of Theoretical Physics, Lanzhou University, Lanzhou, Gansu 730000, China
}
\address{Center for Theoretical Physics of Complex Systems, Institute for Basic Science (IBS), Daejeon 34126, Korea}

\date{\today}

\begin{abstract}
We report on an experimental investigation of the transition of a quantum system with integrable classical dynamics to one with violated time-reversal (\Ti) invariance and chaotic classical counterpart. High-precision experiments are performed with a flat superconducting microwave resonator with circular shape in which \Ti-invariance violation and chaoticity are induced by magnetizing a ferrite disk placed at its center, which above the cutoff frequency of the first transverse-electric mode acts as a random potential. We determine a complete sequence of $\simeq 1000$ eigenfrequencies and find good agreement with analytical predictions for the spectral properties of the Rosenzweig-Porter (RP) model, which interpolates between Poisson statistics expected for typical integrable systems and Gaussian unitary ensemble statistics predicted for chaotic systems with violated \T invariance. Furthermore, we combine the RP model and the Heidelberg approach for quantum-chaotic scattering to construct a random-matrix model for the scattering ($S$) matrix of the corresponding open quantum system and show that it perfectly reproduces the fluctuation properties of the measured $S$ matrix of the microwave resonator.     
\end{abstract}
\bigskip
\maketitle

\section{Introduction\label{Intro}} 
In the past four decades, random matrix theory (RMT)~\cite{Mehta2004} experienced outstanding success in the field of quantum chaos, of which the objective is to identify quantum signatures of classical chaos in the properties of quantum systems. Originally, RMT was introduced by Wigner to describe properties of the eigenstates of complex many-body quantum systems. He was the first to propose that there is a connection between their spectral properties and those of random matrices~\cite{Porter1965,Brody1981,Guhr1989,Weidenmueller2009}. This proposition was taken up in Refs.~\cite{Berry1979,Casati1980,Bohigas1984} and led to the formulation of the Bohigas-Giannoni-Schmit (BGS) conjecture which states that the spectral properties of all quantum systems, that belong to either the orthogonal ($\beta =1$), unitary ($\beta =2$) or symplectic ($\beta =4$) universality class and whose classical analogues are chaotic, agree with those of random matrices from the Gaussian orthogonal ensemble (GOE), the Gaussian unitary ensemble (GUE), or the Gaussian symplectic ensemble (GSE), respectively.  On the other hand, according to the Berry-Tabor (BT) conjecture~\cite{Berry1977}, the fluctuation properties in the eigenvalue sequences of typical integrable systems ($\beta =0$) exhibit Poissonian statistics. 

The BGS conjecture was confirmed theoretically~\cite{LesHouches1989,Haake2018} and experimentally, e.g., with flat, cylindrical microwave resonators~\cite{Sridhar1991,Stein1992,Graef1992,Deus1995,StoeckmannBuch2000}. Below the cutoff frequency $f^{cut}$ of the first transverse-electric mode the associated Helmholtz equation is scalar, that is the electric field strength is parallel to the resonator axis and obeys Dirichlet boundary conditions (BCs) along the side wall. Accordingly, there the Helmholtz equation is mathematically identical to the Schr\"odinger equation of a quantum billiard (QB) of corresponding shape with these BCs and the cavity is referred to as microwave billiard. For generic \Ti-invariant systems with chaotic classical counterpart that are well described by the GOE~\cite{Haake2018}, complete sequences of up to 5000 eigenfrequencies~\cite{Richter1999,Dietz2015a,Dietz2019a} were obtained in high-precision experiments at liquid-helium temperature $T_{\rm LHe}=4$~K with niobium and lead-coated microwave resonators which become superconducting at $T_c=9.2$~K and $T_c=7.2$~K, respectively. The BGS conjecture also applies to quantum systems with chaotic classical dynamics and partially violated \T invariance~\cite{Bohigas1995,Pandey1991,Lenz1992,Guhr1997}. These are described by a HDS model interpolating between the GOE and the GUE for complete \Ti-invariance violation. Such systems were investigated experimentally in~\cite{French1985,mitchell2010,Assmann2016,Pluhar1995} and in microwave billiards~\cite{So1995,Stoffregen1995,Wu1998,Hul2004,Bialous2016,Allgaier2014,Dietz2019b}. In addition, the fluctuation properties of the scattering ($S$) matrix of open quantum systems with partially violated \T invariance were analyzed and exact analytical results were derived based on the Heidelberg approach for quantum-chaotic scattering~\cite{Dietz2007a,Dietz2009,Dietz2010}. Here, \Ti-invariance violation was induced by inserting a ferrite into a microwave billiard with chaotic wave dynamics and magnetizing it with an external magnetic field $B$. Because of the Meissner-Ochsenfeld effect~\cite{Meissner1933} this is not possible at superconducting conditions with a lead-coated cavity~\cite{Dietz2015a} which is a superconductor of type I~\cite{Onnes1911}. To avoid the expulsion of the external magnetic field, the cavity used in~\cite{Dietz2019b} was made from niobium, a type II~\cite{Shubnikov1937} superconductor for 153~mT~$\leq B\leq$~268~mT. 

We report in this work on the experimental investigation of the spectral properties of quantum systems undergoing a transition from Poisson to GUE employing superconducting microwave billiards and the same procedure as in~\cite{Dietz2019b} to induce \Ti-invariace violation. They have the shapes of billiards with integrable dynamics. Ferrites are inserted into the cavities in such a way, that integrability is not destroyed as long as they are not magnetized. We demonstrate that magnetization with an external magnetic field $B$ induces above the cutoff frequency of the ferrite \Ti-invariance violation and also chaoticity. In fact, we showed in Ref.~\cite{Zhang2023a} that above the cutoff frequency the spectral properties of a circular cavity that is loaded with a ferrite material which is magnetized by an external magnetic field perpendicular to the cavity plane, agree with those of a classically chaotic quantum system with a mirror symmetry and completely violated \T invariance. Thus, the magnetized ferrite disk acts like a random potential. The objective is to verify analytical results for the spectral properties of the Rosenzweig-Porter (RP) model~\cite{Rosenzweig1960}, which was intensively studied about 3-4 decades ago~\cite{French1988,Leyvraz1990,Lenz1992,Pandey1995,Brezin1996,Guhr1996a,Guhr1996,Guhr1997,Altland1997,Kunz1998,Frahm1998}. Here, we restrict to the RP model which describes the transition from Poisson to the GUE. We would like to mention that in recent years the RP model has come to the fore in the context of many-body quantum chaos and localization since it undergoes on variation of a parameter $\alpha$ a transition from localized states in the integrable limit via a non-ergodic phase which is characterized by multifractal states to an ergodic phase~\cite{Kravtsov2015,Facoetti2016,Truong2016,Monthus2017,Soosten2019,Pino2019,Tomasi2019,Bogomolny2018,Berkovits2020,Khaymovich2020,Skvortsov2022}. The fractal phase cannot be attained, and above all, not observed with our experimental setup, because we cannot measure wave functions and the achieved values of $\alpha$ are too small. 

Superconductivity of the microwave billiards is crucial in order to obtain complete sequences of eigenfrequencies, however, the construction of such cavities containing niobium parts and the realization of \Ti-invariance violation by magnetizing ferrites positioned inside the cavity with an external field is demanding. Therefore, we first performed experiments with large-scale resonators at room temperature with the sector-shaped cavity shown in the upper part of~\reffig{SketchSM} to test whether we can achieve the transition from Poisson to GUE in such microwave experiments. Due to the large absorption of the ferrites it is not possible to identify complete sequences of eigenfrequencies under such conditions. Yet, another measure for the size of chaoticity and \Ti-invariance violation are the fluctuation properties of the $S$ matrix associated with the resonance spectra of a microwave resonator~\cite{Dietz2009,Dietz2019b}. For the case with no magnetization we analyzed properties of the $S$ matrix of that cavity in Ref.~\cite{Zhang2019} and found clear deviations from RMT predictions. To get insight into the size of chaoticity and \Ti-invariance violation achieved with the cavity with no disks, we compared the fluctuation properties of its $S$ matrix with those of a cavity with a chaotic wave dynamics which was realized by just adding metallic disks as illustrated in the lower part of~\reffig{SketchSM}. It is known that, when magnetizing the ferrites, such systems are well described by the Heidelberg approach for the $S$ matrix~\cite{Mahaux1969} of quantum systems that undergo a transition from GOE to GUE~\cite{Dietz2007,Dietz2009,Dietz2010}. Another objective of these experiments was to compare the properties of the $S$-matrix with a RMT model which we constructed by combining the RP model and the Heidelberg approach. In~\refsec{RT} we report on the scattering experiments and this RMT model and that describing the transition from GOE to GUE. Then, in~\refsec{HeT} we present results for the spectral properties of a superconducting circular cavity containing a ferrite disk at the center which was magnetized with an external magnetic field. These experiments were performed at liquid Helium $T_{\rm LHe}=4$~K. Finally, in~\refsec{Concl} we discuss the results.

\section{Experiments at room temperature\label{RT}}
\subsection{Experimental setup\label{RTExp}}
\begin{figure}[!th]
\includegraphics[width=0.6\linewidth]{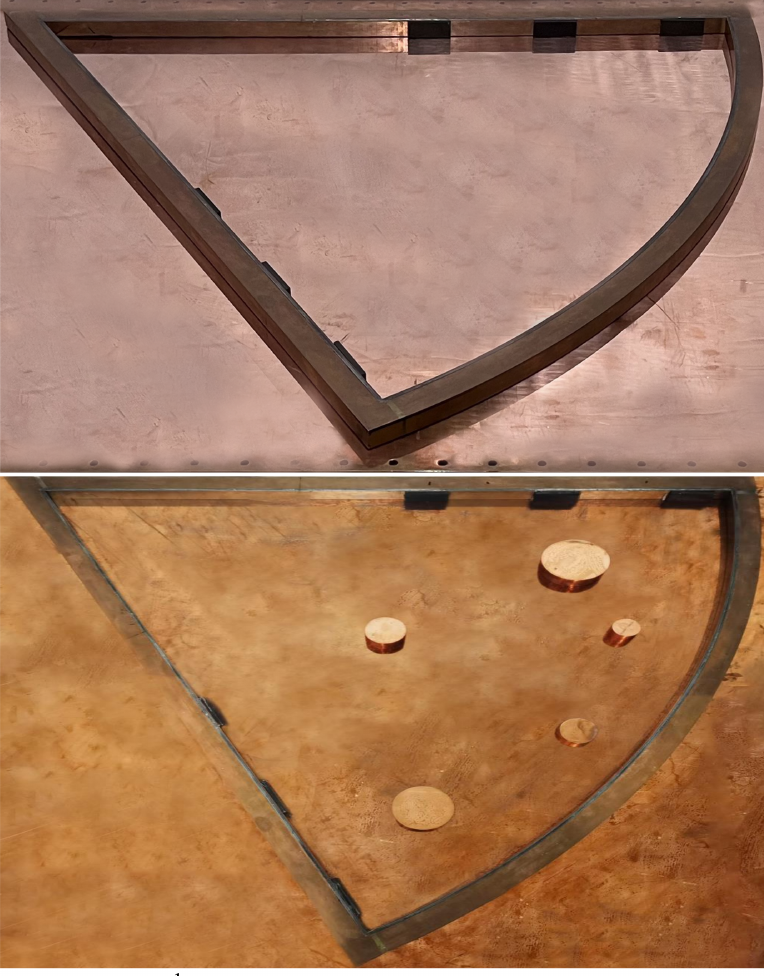}
	\caption{Top: Cavity {\bf SB1} without lid. The microwave resonator is composed of a sector shaped frame, shown in the photograph on top of a plate, which forms its bottom, and a top plate as lid. Bottom: Cavity {\bf SB2} without lid. It is obtained from {\bf SB1} by adding five copper disks of varying sizes and same height as the cavity. All parts are made from copper. To induce partial \Ti-invariance violation, three flat rectangular ferrite pieces (black rectangles) of length 50~mm, width 5~mm and height 20~mm were attached symmetrically to both straight parts of the frame at a distance 425~mm, 575~mm, and 725~mm from the apex and magnetized with an external magnetic field of strength 169~mT. }
\label{SketchSM}
\end{figure}
\begin{figure}[htbp]
        \includegraphics[width=0.8\linewidth]{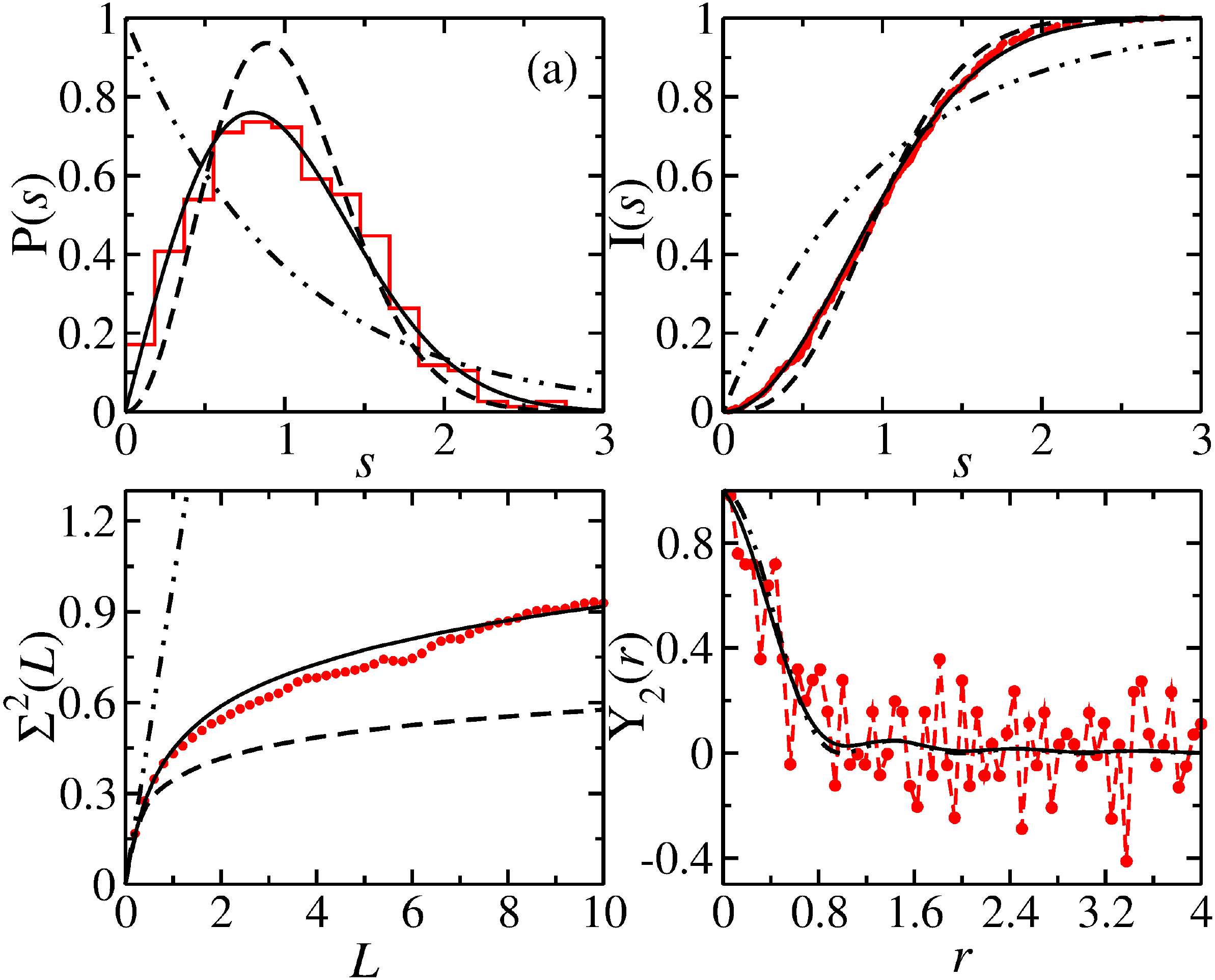}
	\caption{Nearest-neighbor spacing distribution $P(s)$, cumulative nearest-neighbor spacing distribution $I(s)$, number variance $\Sigma^2(L)$ and two-point cluster function $Y_2(r)$ for the cavity {\bf SB2} in the range below the cutoff frequency, which comprises $\approx 500$ eigenfrequencies. They agree well with GOE statistics. The solid, dashed and dashed-dot-dot black lines show the curves for GOE, GUE and Poisson statistics, respectively. For the latter the two-point cluster function equals zero. Therefore it is not shown.}
        \label{Spectr_Sect}
\end{figure}

The room-temperature measurements of the $S$ matrix were performed with the large-scale microwave cavity with the shape of a $60^\circ$ circle sector used in Ref.~\cite{Zhang2019} and shown without lid in the upper part of~\reffig{SketchSM}. We refer to it as {\bf SB1} in the following. The size of the rectangular top and bottom plates are $1260\times 860\times 5$~mm$^3$. The frame has the shape of a $60^\circ$ circle sector with radius $R=800$~mm and height $20$~mm corresponding to a cutoff frequency $7.5$~GHz of the first transverse-electric mode. The top and bottom plate and frame were squeezed together tightly with screw clamps. Furthermore, a rectangular frame of the same size as the plates and the same height as the sector frame, and the top and bottom plates were firmly screwed together. Both frames contained grooves that were filled with a tin-lead alloy to improve the electrical contact. Six thin ferrite strips made of 18F6 with a saturation magnetization $M_{s}=180$~mT were attached symmetrically to the straight side walls of the circle sector, to ensure integrability of the wave dynamics for zero external magnetic field~\cite{Weaver1989,Ellegaard1995,Deus1995,Alt1996,Alt1997,Dembowski2002}. In addition, we inserted five copper disks of varying sizes and same height $20$~mm into the cavity {\bf SB1}, to induce a chaotic dynamics. We refer to it as {\bf SB2} in the following. A photograph of {\bf SB2} without lid is shown in the bottom part of~\reffig{Spectr_Sect}. 

We checked experimentally for frequencies below $f^{cut}=7.5$~GHz that for the case of non-magnetized ferrites the spectral properties of {\bf SB1}~\cite{Berry2008,Bogomolny2009} and {\bf SB2} agree with those of a QB whose shape generates an integrable and chaotic dynamics~\cite{Sinai1970,Bunimovich1979,Berry1981}, respectively. The spectral properties of {\bf SB2}, shown in~\reffig{Spectr_Sect}, agree well with GOE statistics. Those of the empty sector cavity were investigated in~\cite{Zhang2019}. For more details see Secs.~\ref{RMTSpectr} and~\ref{HeTAnal}. Note, that at the walls of the ferrite strips the electric-field strength obeys mixed Dirichlet-Neumann BCs, however, as demonstrated in Refs.~\cite{Berry2008,Bogomolny2009}, the spectral properties of such QBs comply with those of quantum systems with an integrable classical dynamics. 

The eigenfrequencies of the cavity correspond to the positions of the resonances in its reflection and transmission spectra. These were measured by attaching antennas $a$ and $b$ at two out of five possible ports distributed over the cavity lid and connecting them to a Keysight N5227A Vector Network Analyzer (VNA) via SUCOFLEX126EA/11PC35/1PC35 coaxial cables. It couples microwaves into the resonator via one antenna $a$ and receives them at the same or the other antenna $b$, and determines the relative amplitude and phase of the output and input signal, yielding the $S$-matrix elements $S_{aa}$ and $S_{ba}$, respectively. To achieve partial \Ti-invariance violation we magnetized the ferrite pieces with an external magnetic field $B=169$~mT perpendicular to the cavity plane, generated with NdFeB magnets that were placed above and below the cavity~\cite{So1995,Wu1998,Schanze2001,Dietz2007}. The absorption at the ferrite surface is especially high for $B\ne 0$ and Ohmic losses in the walls lead to overlapping resonances, which makes the identifacation of their positions challenging. Above all, the wave dynamics is (nearly) integrable implicating close-lying resonances. As a consequence, we were not able to obtain complete sequences of eigenvalues for the normal-conducting resonators with $B\ne 0$~mT. Yet, we succeeded in constructing a superconducting cavity with induced \T invariance, as outlined in~\refsec{HeT} and, therefore, focused in the room temperature experiments on the fluctuation properties of the $S$ matrix instead. Note that, because we are only interested in properties of the $S$ matrix, we do not need to restrict to the frequency range below $f^{cut}$.

\subsection{Random-matrix formalism for the scattering matrix of a quantum-chaotic scattering process\label{RMTSmat}}
We demonstrated in~\cite{Zhang2019}, that the fluctuation properties of the $S$ matrix of the cavity with no ferrites and no or up to three disks, whose spectral properties follow Poisson and intermediate statistics~\cite{Bogomolny1999}, respectively, clearly deviate from those of chaotic scattering systems. Therefore, the question arose how they look like for cavity {\bf SB1} for $B\ne 0$. In order to get an estimate for the closeness to a chaotic wave dynamics and the size of \Ti-invariance violation we compared its $S$-matrix fluctuation properties to those of the chaotic cavity {\bf SB2}. For the RMT model describing the fluctuation properties of the $S$ matrix of such cavities exact analytical results exist for the two-point $S$-matrix correlation functions for no or partial up to full \Ti-invariance violation in terms of a parameter $\xi$ that quantifies the strength of \Ti-invariance violation. We employ them to estimate it for {\bf SB1} and determine it for {\bf SB2}, as outlined in the following. 

We performed Monte-Carlo simulations based on the scattering formalism for quantum-chaotic scattering~\cite{Mahaux1969}. The $S$-matrix elements of the RMT model, referred to as HDS model in the following,  
\be
S^{HDS}_{ba}(f) = \delta_{ba} - 2\pi i[\hat W^\dagger\left(f\II-\hat H+i\pi\hat W\hat W^\dagger\right)^{-1}\hat W]_{ba}\label{Mahaux}.
\ee
Here, the matrix $\hat W$ accounts for the interaction between the internal states of the resonator Hamiltonian $\hat H$, which mimicks the spectral fluctuation properties of the closed microwave cavity, and the open channels. These comprise the two antenna channels $a$, $b$ and $\Lambda$ fictitious ones that account for Ohmic losses in the walls of the resonator~\cite{Dietz2009,Dietz2010} in terms of a parameter $\tau_{abs}$. The matrix elements of $\hat W$ are real, Gaussian distributed with $W_{a \mu}$ and $W_{b \mu}$ describing the coupling of the antenna channels $a,b$ to the resonator modes $\mu$. We ensured that, as assumed in the HDS model, direct transmission between the antennas is negligible, that is, that the frequency-averaged $S$-matrix is diagonal~\cite{Verbaarschot1985},  implying that $\sum_{\mu = 1}^N W_{e \mu} W_{e^\prime \mu}=N v_{e}^2 \delta_{ee^\prime}$~\cite{Verbaarschot1985}. The parameters $v^2_{e}$ denote the average strength of the coupling of the resonances to channels $e$. For $e=1,\ 2$, referring to the antenna channels, they correspond to the average size of the electric field at the position of the antennas $a$ and $b$ and they yield the transmission coefficients $T_{e} = 1 - \vert\left\langle{S_{ee}}\right\rangle\vert^2$, which are experimentally accessible~\cite{Dietz2010}. The transmission coefficients of the ficitious channels, which are assumed to be the same, $T_3=T_4\dots=T_\Lambda=T_f$, yield through the Weisskopf formula~\cite{Blatt1952} the absorption parameter $\tau_{abs}=\Lambda T_f$. In the numerical simulations we chose $\Lambda =30$.

The transmission coefficients $T_1,T_2$ and $\tau_{abs}$ are input parameters of the HDS model where they are assumed to be frequency-independent. Accordingly, we analyzed the fluctuation properties of the measured $S$ matrix in 1~GHz windows~\cite{Dietz2009}. To determine $\tau_{abs}$ we compared the experimental two-point correlation function
\be
C_{ab}(\varepsilon)=\langle S^{\rm fl}_{ab}(\nu)S^{\ast\rm fl}_{ab}(\nu+\varepsilon)\rangle
\label{Ccorr}
\ee
with $\nu$ and $\varepsilon$ denoting the microwave frequency and frequency-increment in units of the average resonance spacing, to
 RMT predictions. Generally, $\langle\cdot\rangle$ denotes ensemble and spectral averaging. To get an estimate for the strength of \Ti-invariance violation we analyzed cross-correlation coefficients defined as
\be
\label{Ccross}
	C^{cross}_{ab}(0)=\frac{\Re [\langle S^{\rm fl}_{ab}(f)S^{\ast\rm fl}_{ba}(f)\rangle]}{\sqrt{\langle\vert S^{\rm fl}_{ab}(f)\vert^2\rangle\langle\vert S^{\rm fl}_{ba}(f)\vert^2\rangle}},
\ee
which provides a measure for the size of violation of the principle of reciprocity,  $S_{ab}(f)=S_{ba}(f)$, and thus for the strength to \Ti invariance violation. Namely, for \Ti-invariant systems the principle of reciprocity holds and $C^{cross}_{ab}(0)=1$, whereas fully violated \T invariance yields $C^{cross}_{ab}(0)=0$.

We constructed a HDS model for the $S$ matrix of cavity {\bf SB1} by inserting for $\hat H$ in~\refeq{Mahaux} the Hamiltonian of the RP model which describes the transition from Poisson to the GUE,
\be
\label{RPH} \hat H^{0\to 2}(\lambda=\alpha_N/D_N) =\hat H_0+\alpha_N\hat H^{GUE}\, . 
\ee
Here $\hat H_0$ is a random diagonal matrix with a smooth but otherwise arbitrary distribution and $\hat H^{GUE}$ is drawn from the GUE~\cite{Dietz2016}. To get rid of the $N$-dependence of the parameter $\alpha_N$ and to render the limit $N\to\infty$ feasible, which is needed for the derivation of universal, system-independent analytical results for the spectral properties~\cite{Mehta1990,Guhr1998}, the parameter $\alpha_N$ is rescaled with the spectral density of the entries of $\hat H_0$~\cite{Pandey1981,Pandey1995,Guhr1997,Kunz1998,Guhr1998}. For this it is replaced by $\lambda=\alpha_N /D_N$, where $\lambda$ gives the value of $\alpha_N$ in units of $D_N=W/N$, with $W$ denoting the band width of the elements of the diagonal matrix $H_0$. The Hamiltonian  $\hat H^{0\to 2}(\alpha_N)$ interpolates between Poisson for $\alpha_N =0$ and GUE for $\alpha_N\to\infty$, however, its spectral properties already coincide with GUE statistics for values of $\lambda$ of order unity. In the numerical simulations we chose $(400\times 400)$ dimensional random matrices with variances $\langle\Re \left(H^{GUE}_{ij}\right)^2\rangle=\Im\langle\left( H^{GUE}_{ij}\right)^2\rangle=\frac{1}{4N}(1+\delta_{ij})$ and for $\hat H_0$ Gaussian distributed elements with the same variance as for the diagonal elements of $\hat H^{GUE}$. Then, the band width equals $W=2\pi$.

The $S$ matrix properties of cavity {\bf SB2} were compared to Monte-Carlo and analytical results for the $S$ model describing the transition from GOE to GUE~\cite{Dietz2008,Dietz2009,Dietz2010}. For this case $\hat H$ in~\refeq{Mahaux} is replaced by the Hamiltonian~\cite{Pandey1981,Pandey1991,Altland1993} 
\be
\hat H^{1\to 2}(\xi) =\hat H^{(S)}+i\xi\frac{\pi}{\sqrt{N}}\hat H^{(A)},\label{Hxi} 
\ee
with the strength of partial \Ti-invariance determined by the parameter $\xi$. Here, $\hat H^{(S)}$ is a real-symmetric random matrix from the GOE and $\hat H^{(A)}$ is real-antisymmetric one with Gaussian distributed elements with mean-value zero and same variance as for $\hat H^{(S)}$. For the simulations we chose the same values for dimension and variances as for the Hamiltonian in~\refeq{RPH}. 

\subsection{Analysis of the measured scattering matrix\label{RTAnal}}
\begin{figure}[!th]
\includegraphics[width=0.8\linewidth]{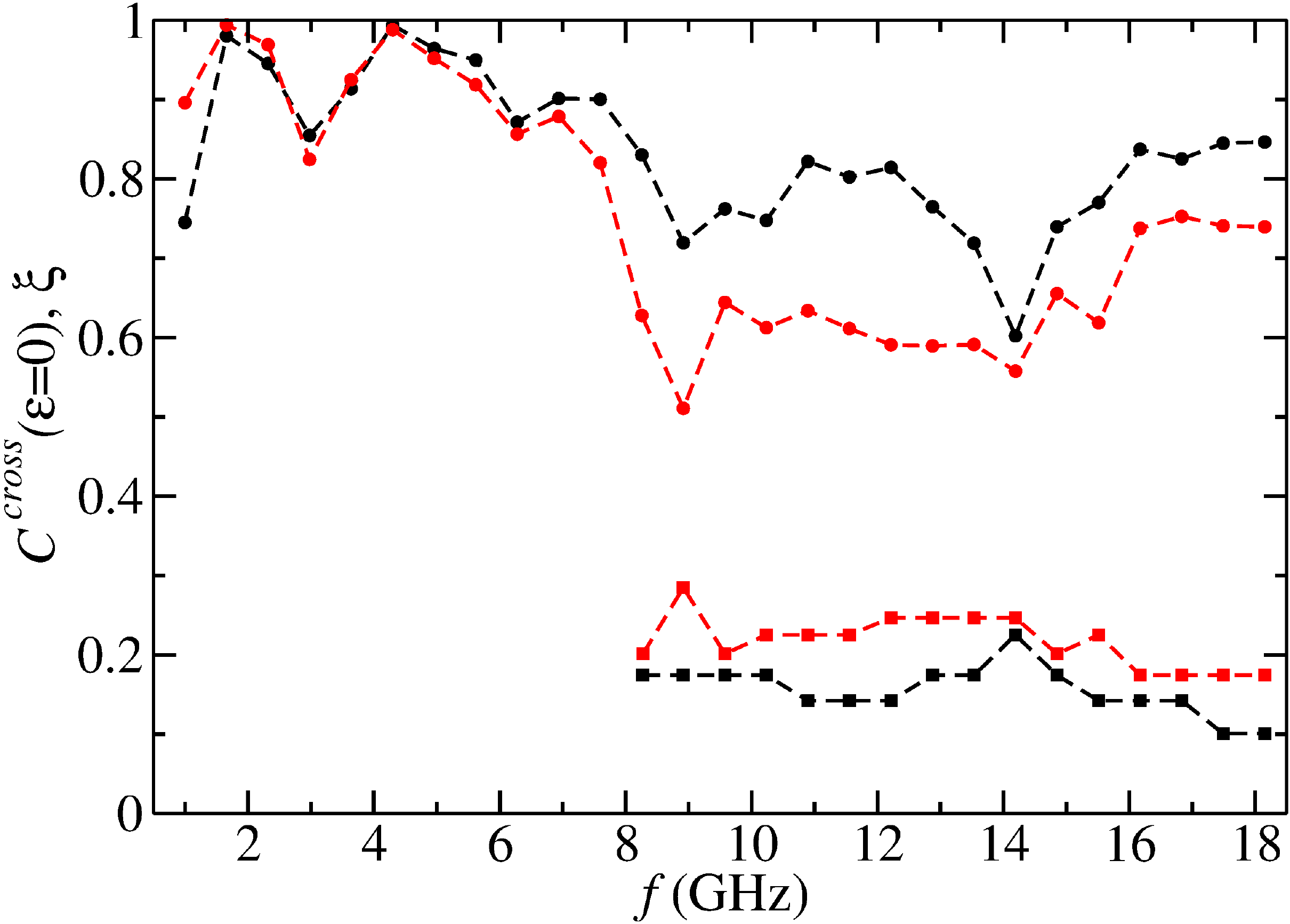}
	\caption{Cross-correlation coefficients of the cavities {\bf SB1} (black dots) and {\bf SB2} (red dots) determined in 1~GHz windows, where the ferrites where magnetized with an external magnetic field with $B=169$~mT. The corresponding values of the \Ti-violation parameter $\xi$ are plotted as black and red squares, respectively (see main text).
        }
\label{CrossCorr}
\end{figure}
\begin{figure}[htbp]
\includegraphics[width=0.8\linewidth]{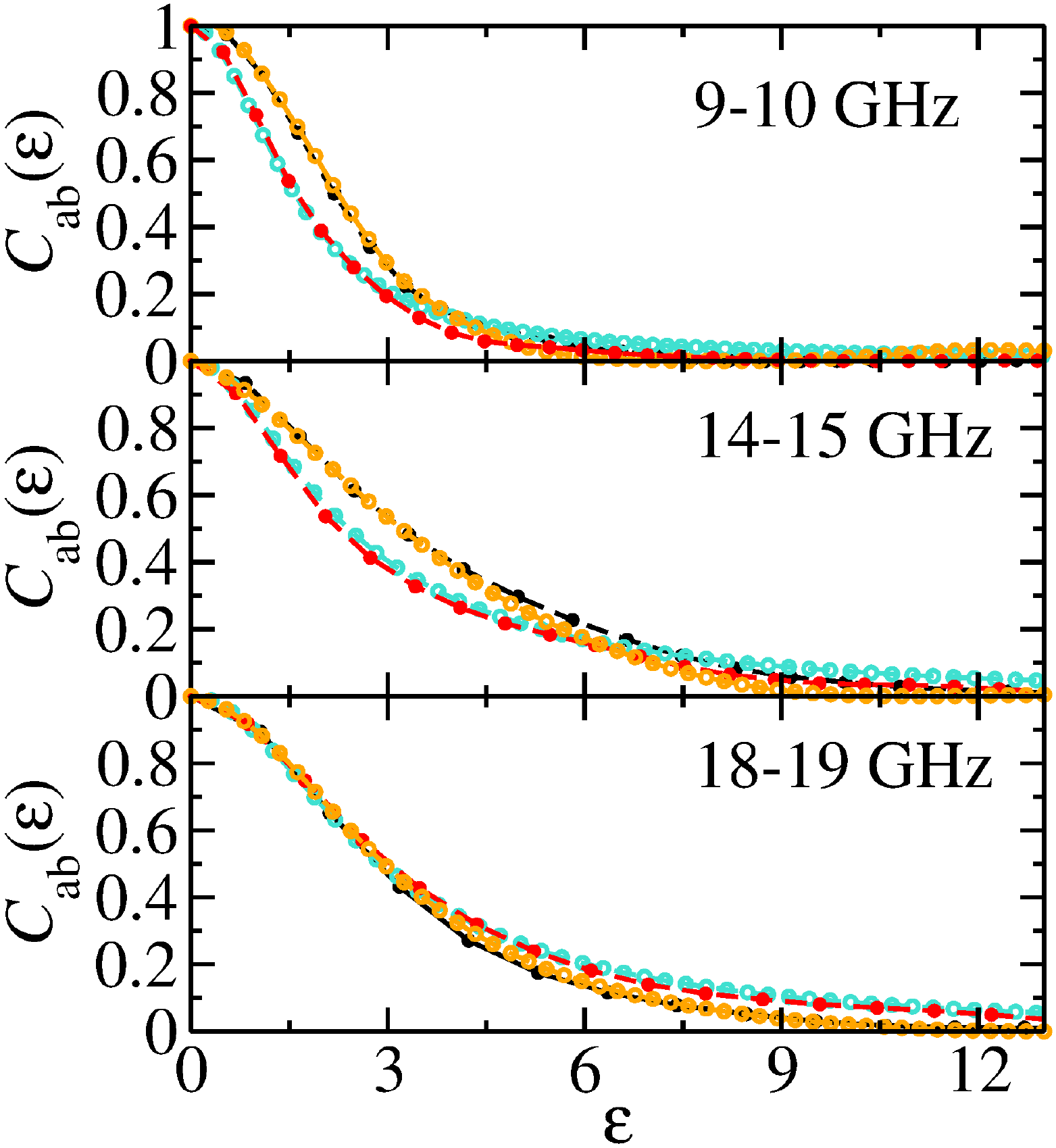} 
	\caption{Two-point correlation functions of the $S$-matrix elements for $a\ne b$ in the frequency ranges indicated in the panels. Shown are the results for the cavity {\bf SB1} (black) and {\bf SB2} (red), the exact analytical results deduced from~\refeq{Mahaux} with $\hat H$ replaced by $\hat H^{1\to 2}(\xi)$ (turquoise) and the results obtained from Monte-Carlo simulations for the $S$ matrix~\refeq{Mahaux} with $\hat H$ replaced by $\hat H^{0\to 2}(\lambda)$ (orange). 
The parameters for the transition GOE to GUE are $T_a=0.60, T_b=0.68, \tau_{abs}=1.6, \xi=0.28$ for $f\in[9,10]$~GHz, $T_a=0.80, T_b=0.87, \tau_{abs}=2.5, \xi=0.2$ for $f\in[14,15]$~GHz and  $T_a=0.86, T_b=0.89, \tau_{abs}=3.75, \xi=0.185$ for $f\in[18,19]$~GHz. For the transition Poisson to GUE they are $T_a=0.60, T_b=0.68,\tau_{abs}=0.75, \lambda =0.25$ for $f\in[9,10]$~GHz, $T_a=0.80, T_b=0.87,\tau_{abs}=2.0, \lambda=0.22$ for $f\in[14,15]$~GHz and  $T_a=0.86, T_b=0.89, \tau_{abs}=3.75, \lambda=0.185$ for $f\in[18,19]$~GHz.}
\label{CorrS}
\end{figure}
\begin{figure}[htbp]
\includegraphics[width=0.8\linewidth]{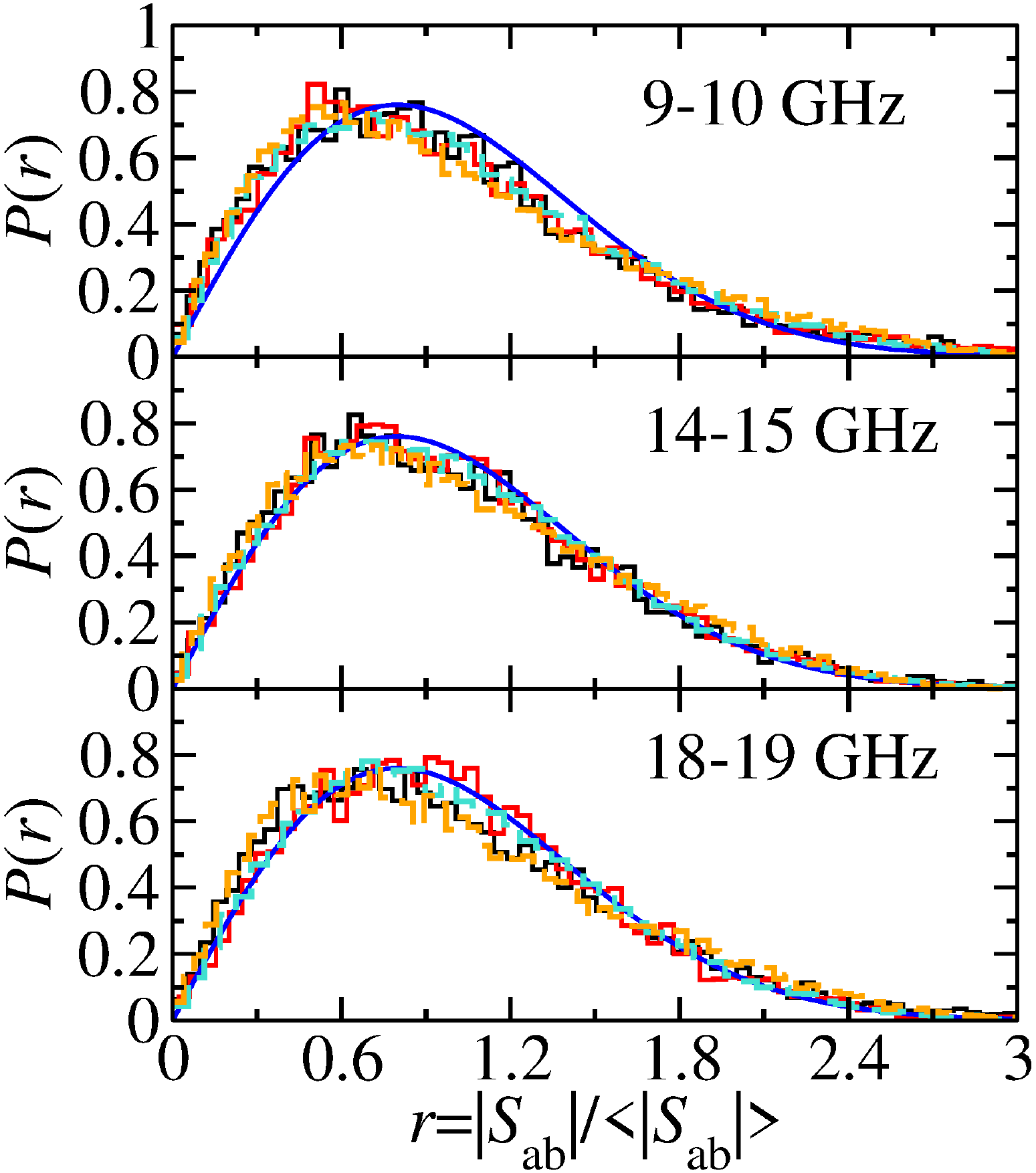}
	\caption{Distributions of the amplitudes of the $S$-matrix for $a\ne b$ in the frequency ranges indicated in the panels. Shown are the results for the cavity {\bf SB1} (black) and {\bf SB2} (red) and the results of Monte-Carlo simulations for the $S$ matrix~\refeq{Mahaux} with $\hat H=\hat H^{1\to 2}(\xi)$ (turquoise) and by $\hat H^{0\to 2}(\lambda)$ (orange) for the same parameters as in~\reffig{CorrS}. The blue solid curve shows the RMT prediction for the Ericson regime of strongly overlapping resonances.}  
\label{VertS}
\end{figure}
To determine the strength $\xi$ of \Ti-invariance violation in the cavity {\bf SB2} we proceeded as in~\cite{Dietz2009,Dietz2010,Bialous2020} and compared the experimental cross-correlation coefficients $C^{cross}_{ab}(0)$, defined in~\refeq{Ccross}, shown in the right part of~\reffig{CrossCorr} as red dots, to exact analytical results for $C^{cross}(0;\xi,T_a,T_b,\tau_{abs})$, yielding the values of $\xi$ shown as red squares in~\reffig{CrossCorr}. As outlined in~\refsec{RMTSpectr}, the cross-correlation coefficient provides a measure for the size of violation of reciprocity, so that we also used it to find out whether \Ti-invariance is violated for the cavity {\bf SB1}. The results are shown as black dots and squares in ~\reffig{CrossCorr}. In order to obtain an estimate for the size of \T invariance violation, we compared the cross-correlation coefficients with the analytical model for the model~\refeq{Hxi}, even though this is not the appropriate model for {\bf SB1}.  Above about 8~GHz \T invariance is clearly violated. 

To determine the value of $\tau_{abs}$ we performed Monte-Carlo simulation for the $S$ matrix~\refeq{Mahaux} with the model~\refeq{RPH}, determined the two-point correlation functions and compared them to the expoermental ones. For the case~\refeq{Hxi} we fit the analytical result for the two-point correlation function to the experimental one. In~\reffig{CorrS} we compare the experimental correlation functions for the cavities {\bf SB1} (black dots) and {\bf SB2} (red dots) for different frequency ranges with the Monte-Carlo and analytical results (orange and turquoise), respectively.  In~\reffig{VertS} are exhibited the corresponding amplitude distributions. For these no analytical results are available for both models. Agreement between the experimental and RMT curves is very good in all cases. We observe, that with increasing frequency the correlation functions for the cavity {\bf SB1} approach those for {\bf SB2}, implying that there a transition from Poisson to GUE takes place. Thus, magnetization of the ferrite pads induces above the $\approx 8$~GHz \Ti-invariance violation and chaoticity of the dynamics, which is above their cutoff frequency. The orange curves, that show the correlation functions and amplitude distributions of the $S$-matrix elements obtained from the HDS model~\refeq{Mahaux} with $\hat H$ replaced by the RP Hamiltonian~\refeq{RPH} agree very well with the experimental ones for the cavity {\bf SB1}. Thus, we may conclude that this HDS model is appropriate for the description of the fluctuation properties of the $S$-matrix of microwave cavities whose wave dynamics undergoes a transition from Poisson to GUE.

\section{Experiments at superconducting conditions\label{HeT}}
\subsection{Experimental setup\label{HeTExp}}
\begin{figure}[!th]
\includegraphics[width=0.8\linewidth]{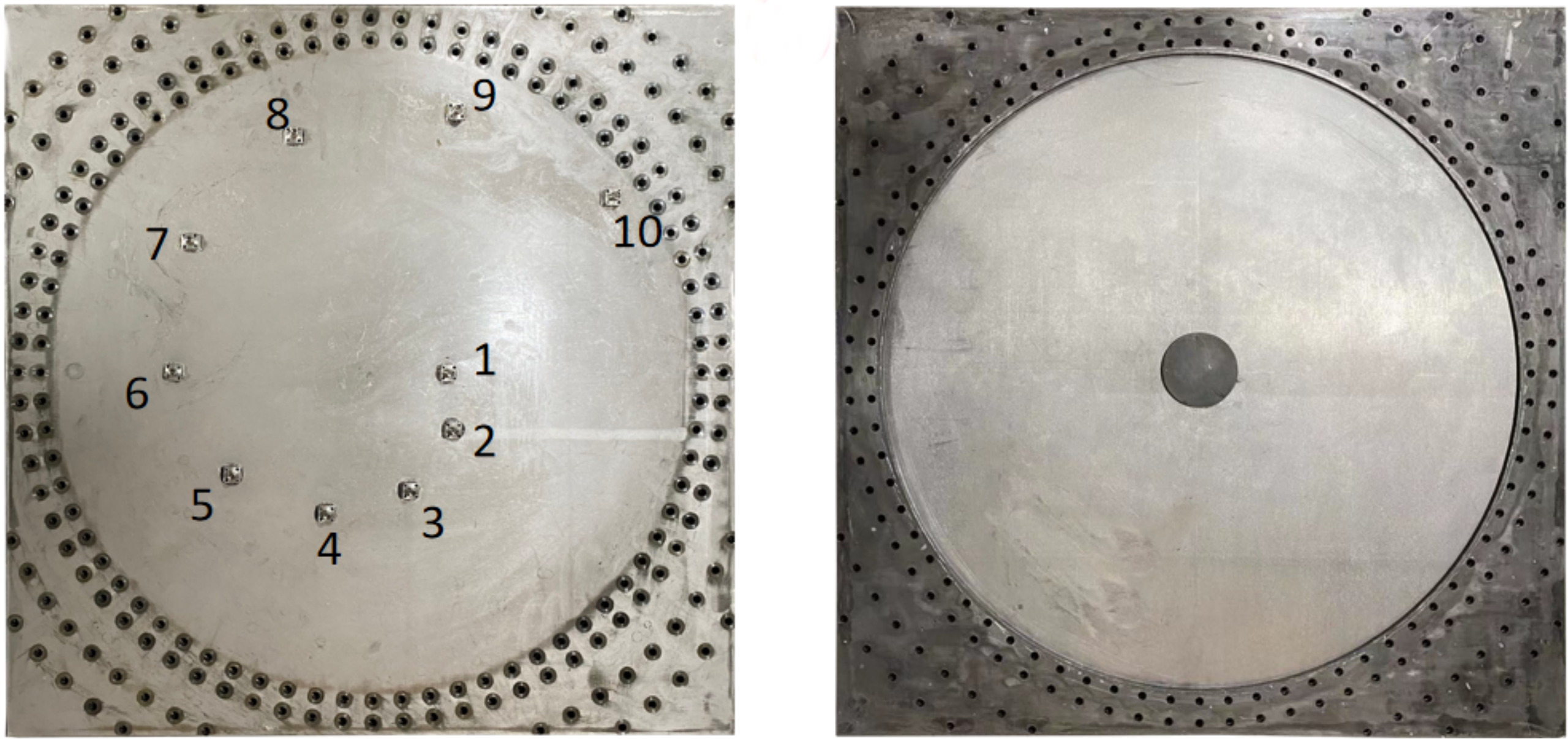}
	\caption{Photograph of the niobium lid (left) and the 5~mm thick lead-coated brass plate with a circular hole on top of a niobium plate (right) of the microwave billiard with a ferrite disk visible at ts center {\bf CB1}. The lid has been removed.}
\label{SketchSC}
\end{figure}
\begin{figure}[!th]
\includegraphics[width=0.8\linewidth]{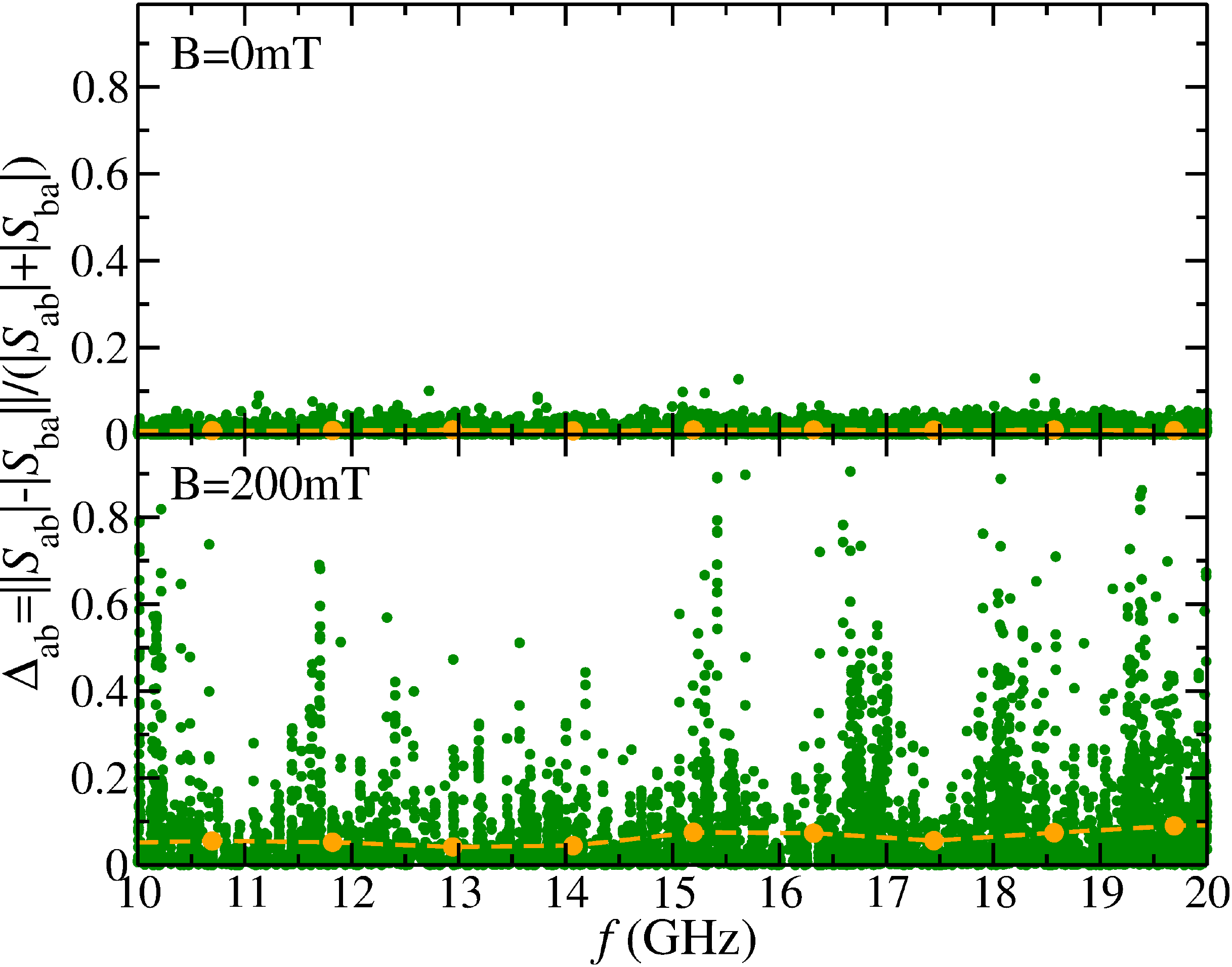}
        \caption{Relative size of violation of the principle of detailed balance for $B=0$~mT and $B=200$~mT, respectively. The orange dots connected by dashed line show the average values in a sliding 1~GHz windows.
        }
\label{DiffSab}
\end{figure} 
We performed experiments at superconducting conditions with a circular microwave billiard, referred to as {\bf CB1} in the following, with radius $R=250$~mm containing a ferrite disk with radius $R_0=30$~mm at the center, shown in~\reffig{SketchSC}, to investigate spectral properties of quantum systems that undergo a transition from integrable classical dynamics with preserved \T invariance to a chaotic one with complete \Ti-invariance violation.  The radius of the circle is $R=250$~mm and the cavity height equals $h=5$~mm corresponding to a cutoff frequency $f^{cut}=30$~GHz. A ferrite disk made of 19G3 with saturation magnetization $M_s=195$~mT with radius $R_0=30$~mm and same height as the cavity corresponding to a cutoff frequency $f_F^{cut}\approx 4.5$~GHz is placed at its center. To induce \Ti-invariance violation the ferrite is magnetized with a static magnetic field of strength $B=200$~mT that is generated with two external NdFeB magnets, fixed above and below the cavity~\cite{Dietz2019b}. In total 10 ports were fixed to the lid. The three plates are screwed together tightly through holes along the cavity boundary and circles visible in the photographs, and tin-lead is filled into grooves that were milled into the top and bottom surfaces of the middle plate along the circle boundary to attain a good electical contact and, thus, high quality factors. To achieve a high quality factors of $Q\gtrsim 5\cdot 10^4$, the cavity was cooled down to below $\approx 5$~K in a cryogenic chamber constructed by ULVAC Cryogenics in Kyoto, Japan. We thereby could determine a complete sequence of 1014 eigenfrequencies in the frequency range 10-20~GHz, using the resonance spectra measured between the antennas for all port combinations. The measurements were performed for $B=0$~mT and $B=200$~mT. We also determined the eigenfrequencies of the circular cavity with a metallic disk instead of the ferrite at the center, denoted  {\bf CB2} in the following. Below $f^{cut}$ it corresponds to an integrable ring-shaped QB. 
\begin{figure}[htbp]
\includegraphics[width=0.8\linewidth]{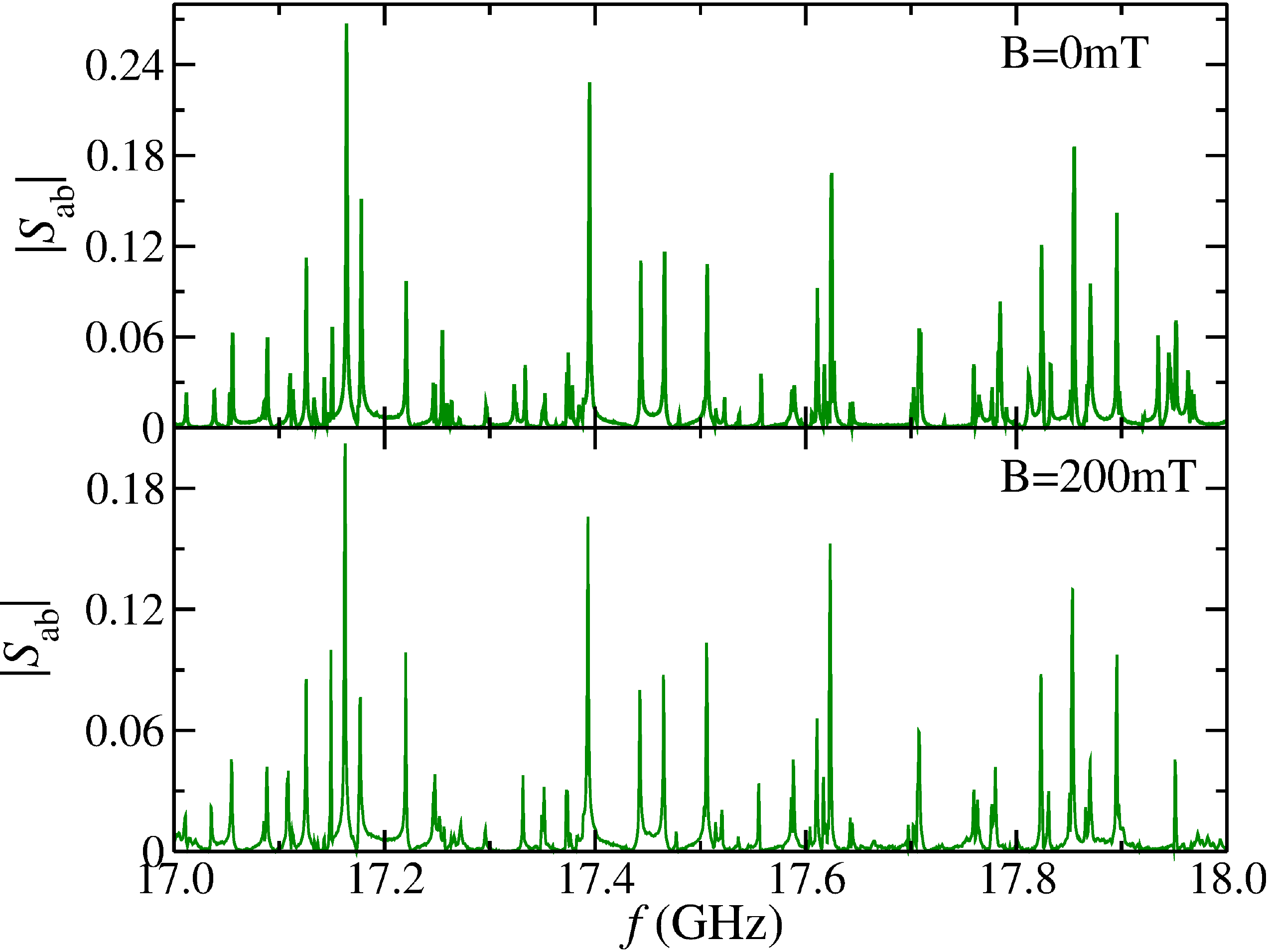}
        \caption{Part of the transmission spectrum of the circular microwave billiard measured at $\approx 5$~K for external magnetic field $B=0$~mT and $B=200$~mT.}
\label{TransmSC}
\end{figure}

In~\reffig{DiffSab} we show $\Delta_{ab}=[\vert\vert S_{ab}\vert-\vert S_{ba}\vert\vert]/[\vert S_{ab}\vert+\vert S_{ba}\vert]$, which gives a measure for the violation of detailed balance, $\vert S_{ab}\vert=\vert S_{ba}\vert$, and thus for the strength of \Ti-invariance violation. Note that for the calibration of the $S$ matrix at superconducting conditions a special cumbersome procedure is required~\cite{Marks1991,Rytting2001,Yeh2013} which, however, is not needed as long as one only is interested in spectral properties. Therefore, we cannot get any information on \Ti-invariance violation from the fluctuation properties of the $S$ matrix, that depend on its phases, like the cross-correlation coefficients, and considered $\Delta_{ab}$ instead~\cite{Dietz2010}. For $B=0$~mT the principle of detailed balance is fulfilled up to experimental accuracy, whereas for $B=200$~mT it is clearly violated. In~\reffig{TransmSC} we compare measured transmission spectra of the cavity with a ferrite disk at the center for $B=0$~mT and $B=200$~mT. The effect of magnetization is that the resonances are shifted with respect to those for $B=0$~mT, which becomes visible in a change of the spectral properties as demonstrated in~\refsec{HeTAnal}, and part of them are missing.

\subsection{Review of analytical results for the RP model\label{RMTSpectr}}
We analyzed the spectral properties in terms of the nearest-neighbor spacing distribution $P(s)$, the cumulative nearest-neighbor spacing distribution $I(s)$, the two-point cluster function $Y_2(r)$, which is related to the spectral two-point correlation function $R_2(r)$ via $Y_2(r)=1-R_2(r)$, the number variance $\Sigma^2 (L)=\langle (N(L)-\langle N(L)\rangle)^2\rangle$, and the form factor $K(\tau)=1-b(\tau)$ with $b(\tau)=\int_{-\infty}^\infty Y_{2}(r)e^{-ir\tau}dr$. We compared these measures to analytical ones for the RP model~\refeq{RPH}. The parameter $\alpha_N$, respectively $\lambda=\alpha_N /D_N$ characterizing the transition from Poisson to GUE was determined by fitting the result for the number variance $\Sigma^2_{0\to 2}(L)$, which was deduced from the exact analytical result for the two-point cluster function $Y_2^{0\to 2}(r)$ derived in Refs.~\cite{Lenz1992,Kunz1998,Frahm1998}. We would like to note, that Georg Lenz derived an analytical expression already in 1992, which is exact for all values of $\alpha_N$ and dimensions $N$ of $\hat H^{0\to 2}(\lambda)$ in~\refeq{RPH}, however, the $N$ dependence is so complex, that the computation of the limit $N\to\infty$ was impossible~\cite{Lenz1992}. In Ref.~\cite{Kunz1998} $Y_2^{0\to 2}(r)$ was obtained from the inverse Fourier transform of the analytical result for $b^{0\to 2}(\tau)=\int_{-\infty}^\infty Y_{2}^{0\to 2}(r)e^{-ir\tau}dr$, 
\ba
&&K^{0\to 2}(\tilde\tau)=1+\frac{2}{\gamma}I_1(\gamma)\exp{\left[-\pi\tilde\alpha^2\tilde\tau-\frac{\tilde\alpha^2\tilde\tau^2}{2}\right]}
\label{B2Anal}\\
&&-\frac{\tilde\tau}{2\pi}\gamma\int_1^\infty dt(t^2-1)I_1(\gamma t)\exp\left[-t^2\frac{\tilde\alpha^2\tilde\tau^2}{2}-\pi\tilde\alpha^2\tilde\tau\right],
\nonumber\\
&&\gamma =\sqrt{2\pi}\tilde\alpha^2\tilde\tau^{3/2},
\ea
which was rederived in Ref.~\cite{Kravtsov2015}. In Ref.\cite{Frahm1998} an exact analytical expression was computed for $Y_2^{0\to 2}(r)$ based on the graded eigenvalue method, yielding
\ba
Y_2^{0\to 2}(r)&&=\frac{1}{2(\pi r)^2}\left[1-e^{-2\frac{r^2}{\tilde\alpha^2}}\cos(2\pi r)\right]-\frac{1}{(\pi\tilde\alpha)^2}
\label{Y2Anal}\\
&&+\frac{1}{\pi}\int_0^\infty rdr e^{-\frac{r^2}{2c}}\int_0^\pi d\phi\cos(\phi)\left[\Re (A)+\Re (B)\right]\nonumber\\
A&&=\frac{e^{i\phi}\left[1-\frac{r}{\kappa}\sin\phi\right]}{1+i\frac{re^{i\phi}}{2\kappa}}\exp\left[-i\frac{r^2}{2c\kappa}\frac{1}{1-\frac{r}{\kappa}\sin\phi}\right]\nonumber\\
B&&=\frac{e^{-i\phi}\left[1+\frac{r}{\kappa}\sin\phi\right]}{1+i\frac{re^{-i\phi}}{2\kappa}}\exp\left[-i\frac{r^2}{2c\kappa}\frac{1}{1+\frac{r}{\kappa}\sin\phi}\right]\nonumber ,\\
\kappa &&=\frac{r}{\pi\tilde\alpha ^2},\, c=\frac{1}{(\pi\tilde\alpha)^2}.\nonumber
\ea
Note, that there are discrepancies in the scales of $\tilde\alpha$ and $\tilde\tau$ between Refs.~\cite{Altland1997,Kunz1998,Guhr1998,Kravtsov2015}, resulting from differeing definitions of the $N$-independent parameter $\lambda$. We fixed the scales and verified the validity of Eqs.~(\ref{B2Anal})-(\ref{Sigma2Anal}) by comparing the analytical results with Monte-Carlo simulations for spectra consisting of a few hundreds of eigenvalues. We chose  $(400\times 400)$-dimensional random matrices, such that the number of eigenvalues is comparable to the experimental eigenfrequency sequences. Here, we used the same settings as in~\refeq{RPH}, that is, Gaussian distributed entries for $\hat H_0$ of the same variance $\langle (\hat H_0)_{ii}\rangle^2=\frac{1}{2N}$ as for the diagonal elements $\hat H^{GUE}$, such that their band width equals $W=2\pi$, that is, $\alpha_N =\lambda D_N=\lambda\frac{2\pi}{N}$ in the random matrix model~\refeq{RPH} with a $N$-dimensional Hamiltonian. This yielded $\tilde\alpha=\frac{\pi}{\sqrt{2}}\lambda$ and $\tilde\tau=\frac{\tau}{\tilde\alpha^2}$. Note, that approximations have been derived for $Y_2^{0\to 2}(r)$ for $\lambda \ll 1$ and $\lambda\gg 1$ in Refs.~\cite{French1988,Leyvraz1990,Lenz1992,Pandey1995,Brezin1996,Guhr1996a,Guhr1996,Guhr1997,Altland1997,Kunz1998,Frahm1998}. These, however, are applicable to the experimental data for a small value of $r$, respectively $L$. Therefore, we do not show the comparison. We determined the values of $\lambda$ from the experimental eigenfrequency spectra by fitting the analytical expression for the number variance deduced from~\refeq{Sigma2Anal} via the relation
\be
\label{Sigma2Anal}
\Sigma_{0\to 2}^2(L)=L-2\int_0^L(L-r)Y_2^{0\to 2}(r)dr\, ,
\ee
to their number variance.

In Ref.~\cite{Lenz1992} a Wigner-surmise like expression was derived for the nearest-neighor spacing distribution based on the RP model~\refeq{RPH} with $N=2$, which was rederived in Ref.~\cite{Kota1999} and is quoted in Ref.~\cite{Schierenberg2012}, given by
\ba
\label{PS}
        &&P_{0\to 2}(s)=Cs^2e^{-D^2s^2}\int_0^\infty dxe^{-\frac{x^2}{4\alpha_L^2}-x}\frac{\sinh z}{z},\,\\
        &&\nonumber D(\alpha_L)=\frac{1}{\sqrt{\pi}}+\frac{1}{2\alpha_L}e^{\alpha_L^2}{\rm erfc}(\alpha_L)-\frac{\alpha_L}{2}{\rm Ei}\left(\alpha_L^2\right)\\
        &&\nonumber +\frac{2\alpha_L^2}{\sqrt{\pi}}{_2{F}_2}\left(\frac{1}{2},1;\frac{3}{2},\frac{3}{2};\alpha_L^2\right),\\
        &&\nonumber C(\alpha_L)=\frac{4D^3(\alpha_L)}{\sqrt{\pi}},\, z=\frac{xDs}{\alpha_L},
\ea
where erfc$(x)$ denotes the complementary error function, Ei$(x)$ the exponential integral, and ${_2{F}_2}(i\alpha_1,\alpha_2;\beta_1,\beta_2;x)$ the generalized hypergeometric error function~\cite{Abramowitz2013,Gradshteyn2007}. Comparison of these analytical results with our Monte-Carlo simulations yields $\alpha_L =\sqrt{2}\lambda$. 

We also analyzed the distribution of the ratios~\cite{Oganesyan2007,Atas2013} of consecutive spacings between next-nearest neighbors, $r_j=\frac{f_{j+1}-f_{j}}{f_{j}-f_{j-1}}$ for which no anayltical results are available. Yet, they have the advantage that no unfolding is requirted since the ratios are dimensionless~\cite{Oganesyan2007,Atas2013,Atas2013a}. Another frequently studied measure is the power spectra, defined as 
\be
s\left(\tau=\frac{l}{n}\right)=\left\langle\left\vert\frac{1}{\sqrt{n}}\sum_{q=0}^{n-1} \delta_q\exp\left(-2\pi i\frac{l}{n} q\right)\right\vert^2\right\rangle,\, l=1,\dots n
\ee
with $n$ denoting the number of eigenvalues and $\delta_q=\epsilon_{q+1}-\epsilon_1-q$ for a complete sequence of $n$ levels, where $\frac{1}{n}\leq\tau\leq 1$~\cite{Relano2002,Faleiro2004}. An analytical expression was derived for $s(\tau)$ in Ref.~\cite{Faleiro2004} in terms of the spectral form factor. It provides a good approximation for experimental data obtained in microwave networks and microwave billiards~\cite{Faleiro2006,Bialous2016,Bialous2016a,Che2021} consisting of sequences of a few hundreds of eigenfrequencies, for all three universality classes of Dyson's threefold way. With the aim to get an approximation for $s(\tau)$ for the transition from Poisson to GUE, we replaced in the analytical expression of Ref.~\cite{Faleiro2004} the spectral form factor by the expression~\refeq{B2Anal}, yet didn't find good agreement with the experimental data, also not in Monte-Carlo simulations with high-dimensional RP Hamiltonians~\cite{Cadez2023}. Exact analytical results were obtained for the power spectra in Refs.~\cite{Riser2017,Riser2023} for fully chaotic quantum systems with violated \T invariance, however, we are not aware of any analytical results for the RP model. Due to the lack of an analytical expression we compared the results deduced from the experimental data for the power spectrum to Monte-Carlo simulations, as outlined in~\refsec{HeTAnal}. 

\subsection{Analysis of correlations in the eigenfrequency spectra\label{HeTAnal}}
For the analysis of the spectral properties we unfolded the eigenfrequencies to mean spacing one, by replacing them by the spectral average of the integrated resonance density $\langle\mathcal{N}(f)\rangle$, $\epsilon_i=\langle\mathcal{N}(f_i)\rangle$, which for the cavity {\bf CB2} is given by Weyl's formula, $\langle\mathcal{N}(f)\rangle=\frac{\mathcal{A}\pi}{c^2}f^2-\frac{\mathcal{L}}{2c}f+N_0$, with $\mathcal{A}$ and $\mathcal{L}$ denoting the area and perimeter of the billiard, respectively, and provides a good approximation for the cavity {\bf CB1} for $B=0$~mT. For $B\ne 0$~mT we determined $\langle\mathcal{N}(f)\rangle$ by fitting a quadratic polynomial to the experimentally determined $\mathcal{N}(f)$. The parameter $\lambda$ was determined in all considered cases by fitting the analytical result~\refeq{Sigma2Anal} to the experimentally determined number variance, which provides a suitable measure since it is very sensitive to small changes in $\lambda$.

\begin{figure}[ht]
\includegraphics[width=0.8\linewidth]{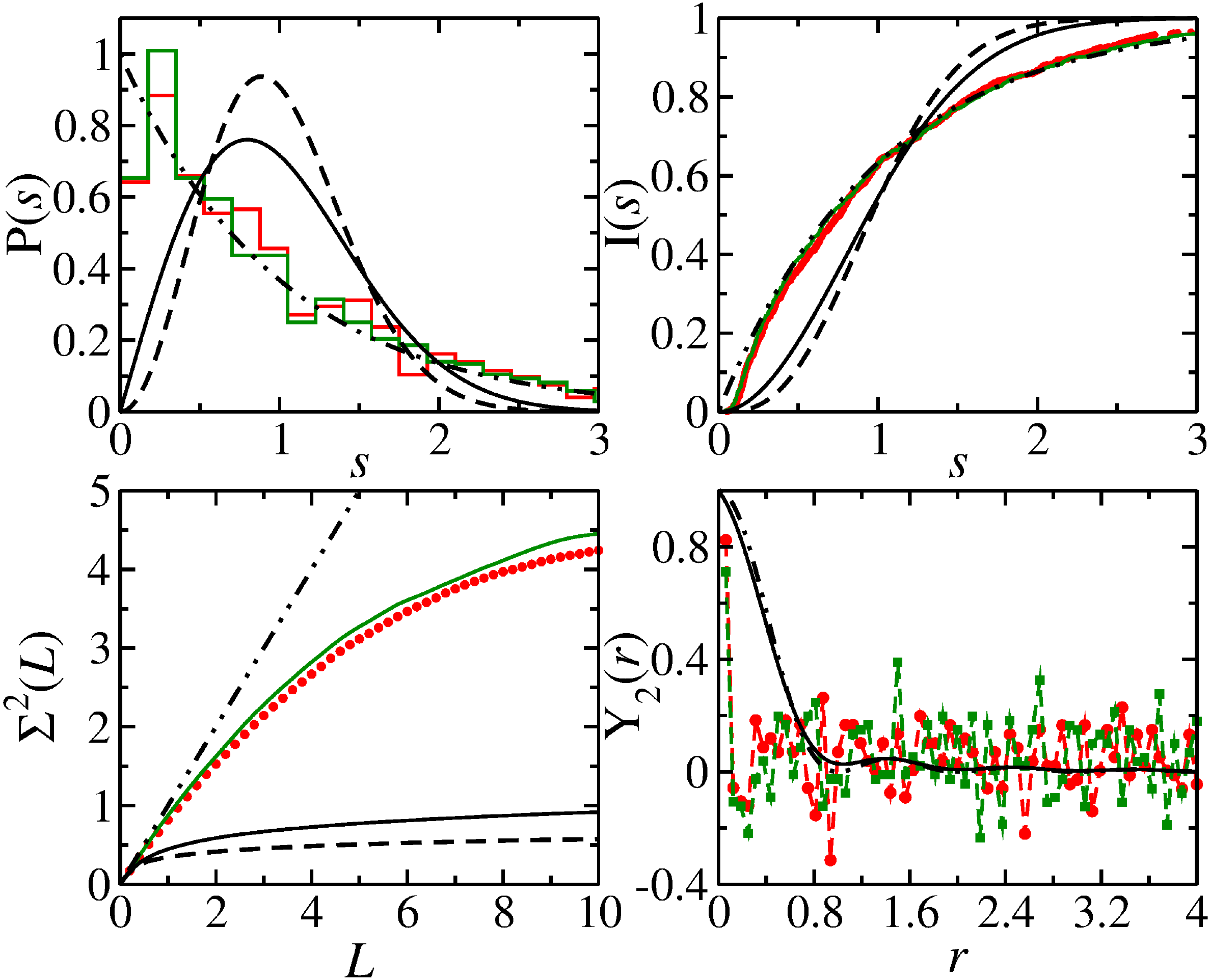}
\includegraphics[width=0.8\linewidth]{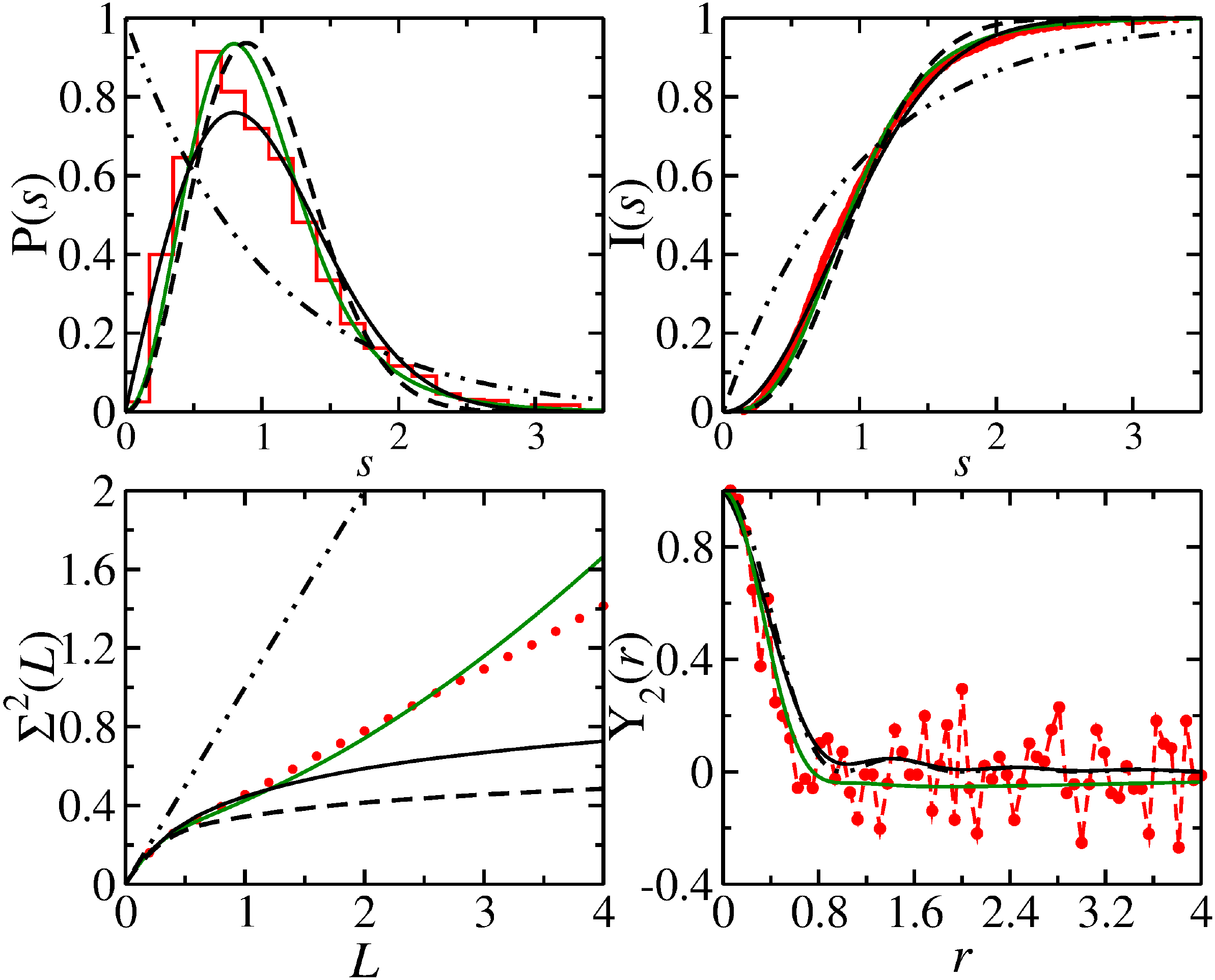}
	\caption{Top: Spectral properties of the cavities {\bf CB1} with $B=0$~mT (red histogram and dots) and {\bf CB1} (green histogram and squares) for $\approx 1000$ eigenfrequencies in the frequency range [10,20]~GHz. They are compared to the results for Poisson (black dashed-dot-dot lines), GOE (solid black lines) and GUE (dashed black lines) statistics.
	Bottom: Same as top for the cavity {\bf CB1} with $B=200$~mT. Here, the green lines show the curves deduced from Eqs.~(\ref{Y2Anal}),~(\ref{Sigma2Anal}) and~(\ref{PS}) for $\lambda=0.475$.}
\label{SpectralAll}
\end{figure}
\begin{figure}[ht]
\includegraphics[width=0.8\linewidth]{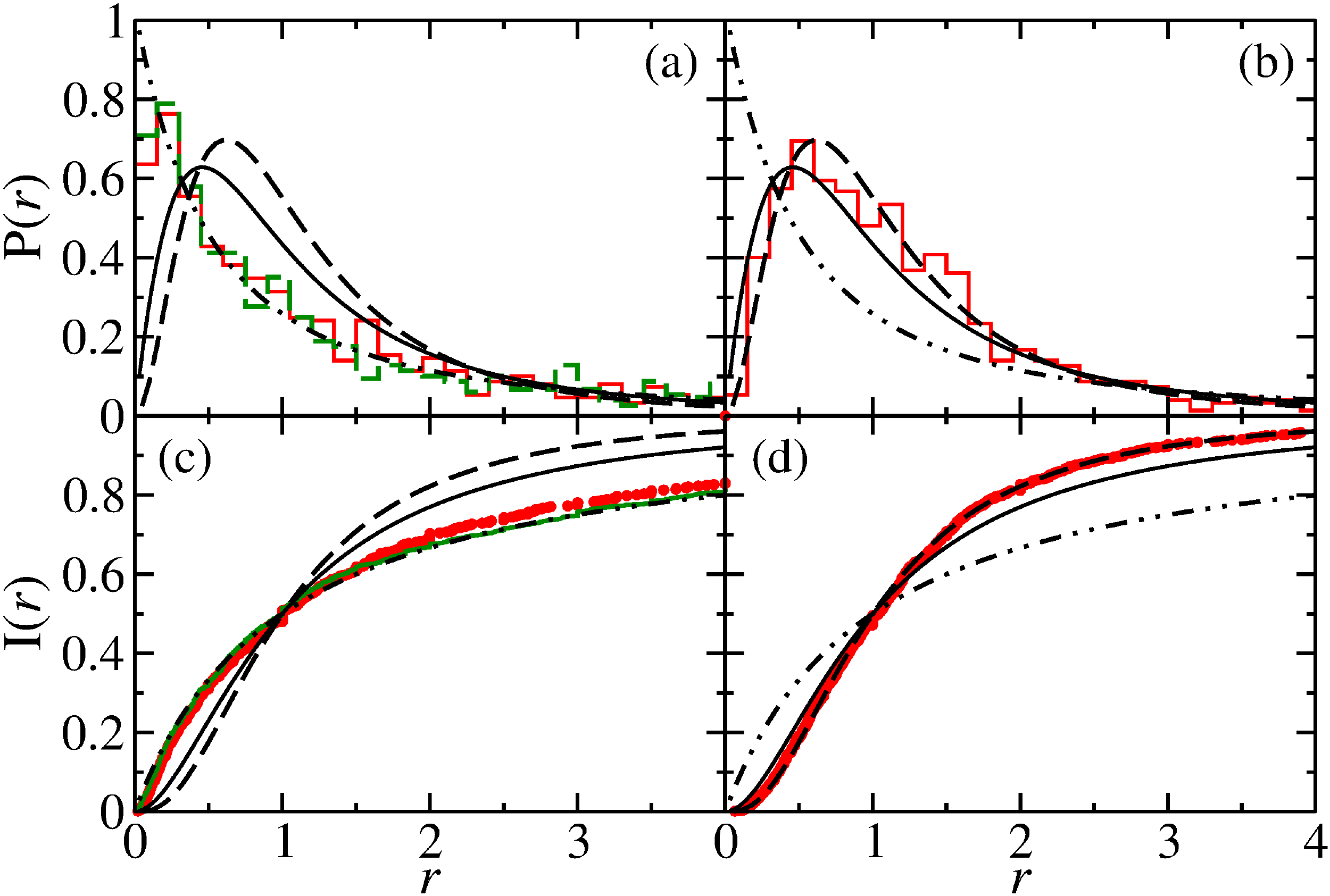}
	\caption{Ratio distributions (upper panel) and cumulative ratio distributions (lower panels). (a), (c):  Cavities {\bf CB1} (red histogram and dots) amd {\bf CB2} (green histogram and squares) for $\approx 1000$ eigenfrequencies in the frequency range [10,20]~GHz.  They are compared to the results for Poisson (black dashed-dot-dot lines), GOE (solid black line) and GUE (dashed black lines) statistics. (b), (d): Same as (a), (c) for $B=200$~mT.}
        \label{Ratiod}
\end{figure}

In the upper part of~\reffig{SpectralAll} we show spectral properties of the cavities {\bf CB1} with $B=0$~mT and {\bf CB2}. The curves lie very close to or on top of each other and coincide with analytical results for the corresponding ring QB, that is, the agreement with Poisson is as good as expected for $\approx 1000$ levels. In the lower part are exhibited the spectral properties for the cavity {\bf CB1} with $B=200$~mT in the range $[0,20]$~GHz, which also comprises $\approx 1000$ levels. They agree best with the RP model for $\lambda=0.474$. In~\reffig{Ratiod} are shown the associated ratio distributions. They are close to Poisson for the case $B=0$~mT and to GUE for $B=200$~mT. 

\begin{figure}[ht!]
        \includegraphics[width=0.65\linewidth]{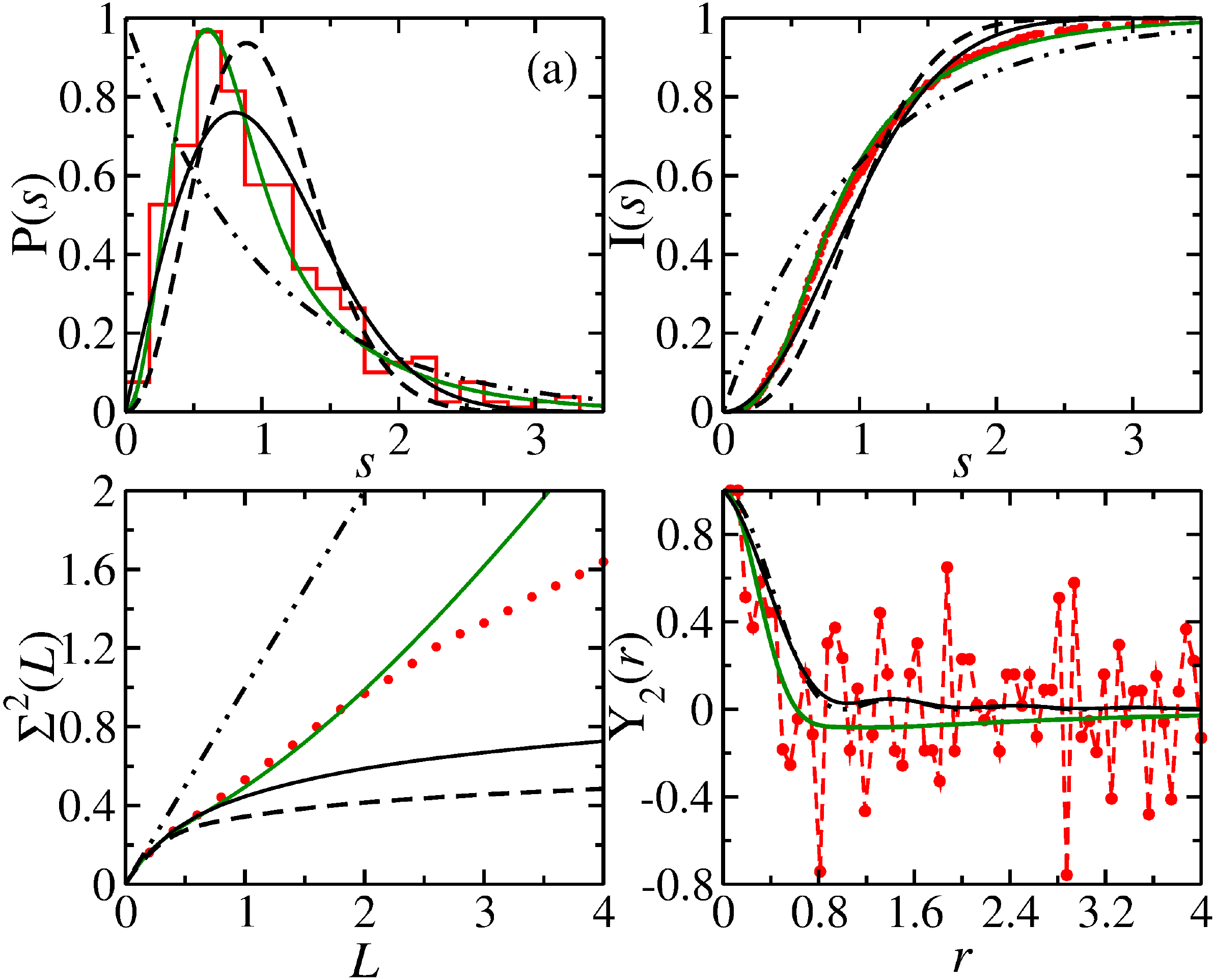}
        \includegraphics[width=0.65\linewidth]{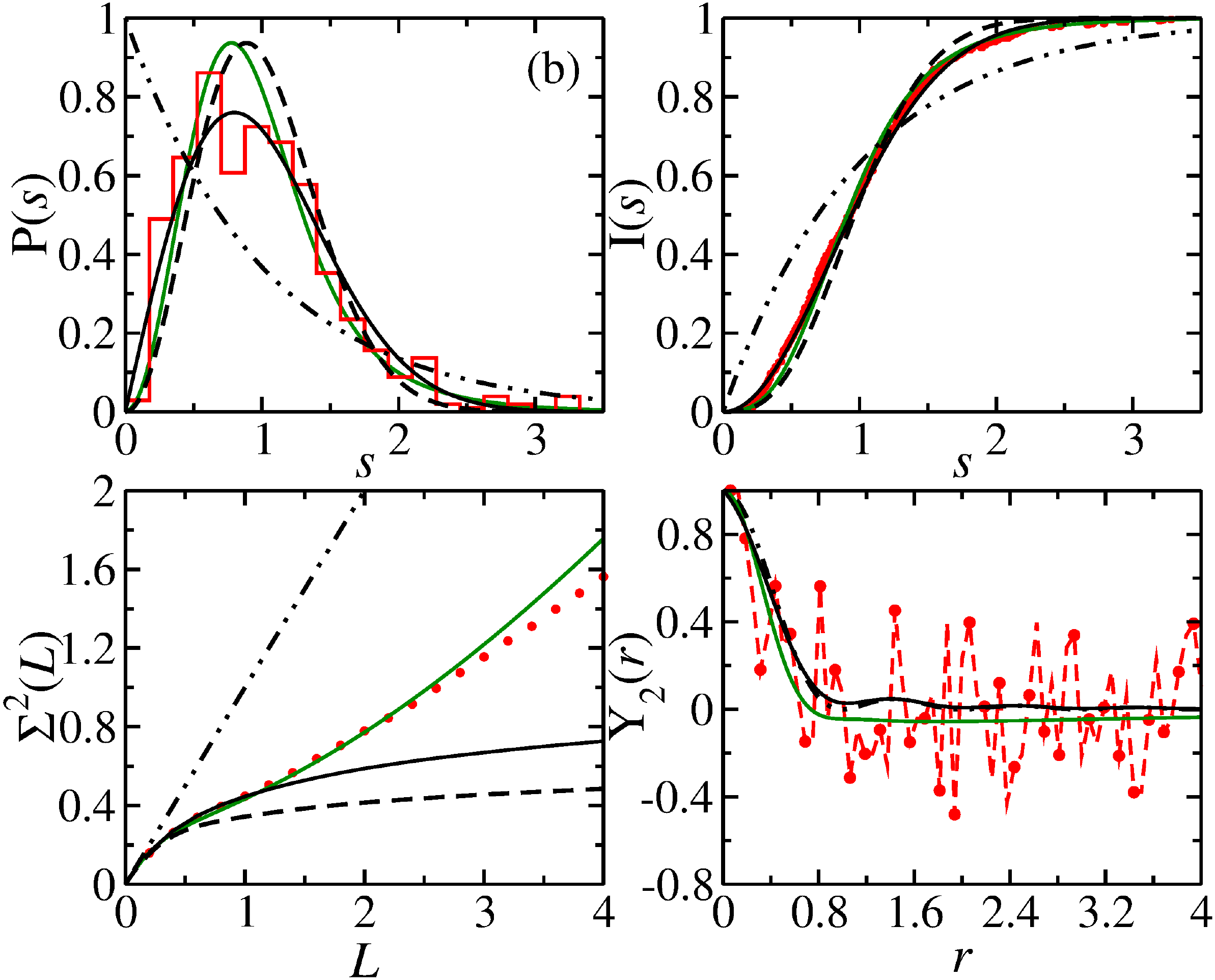}
        \includegraphics[width=0.65\linewidth]{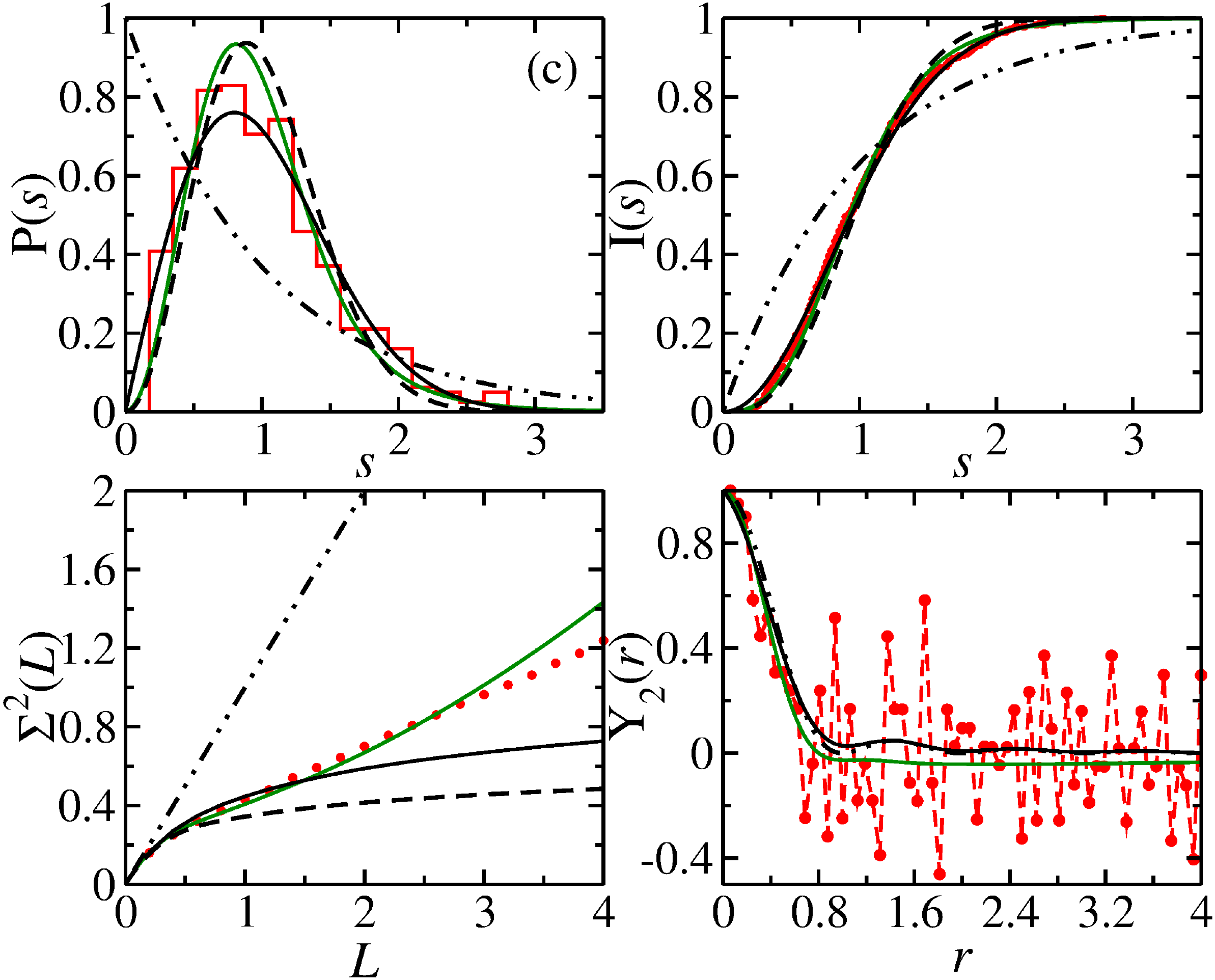}
        \includegraphics[width=0.65\linewidth]{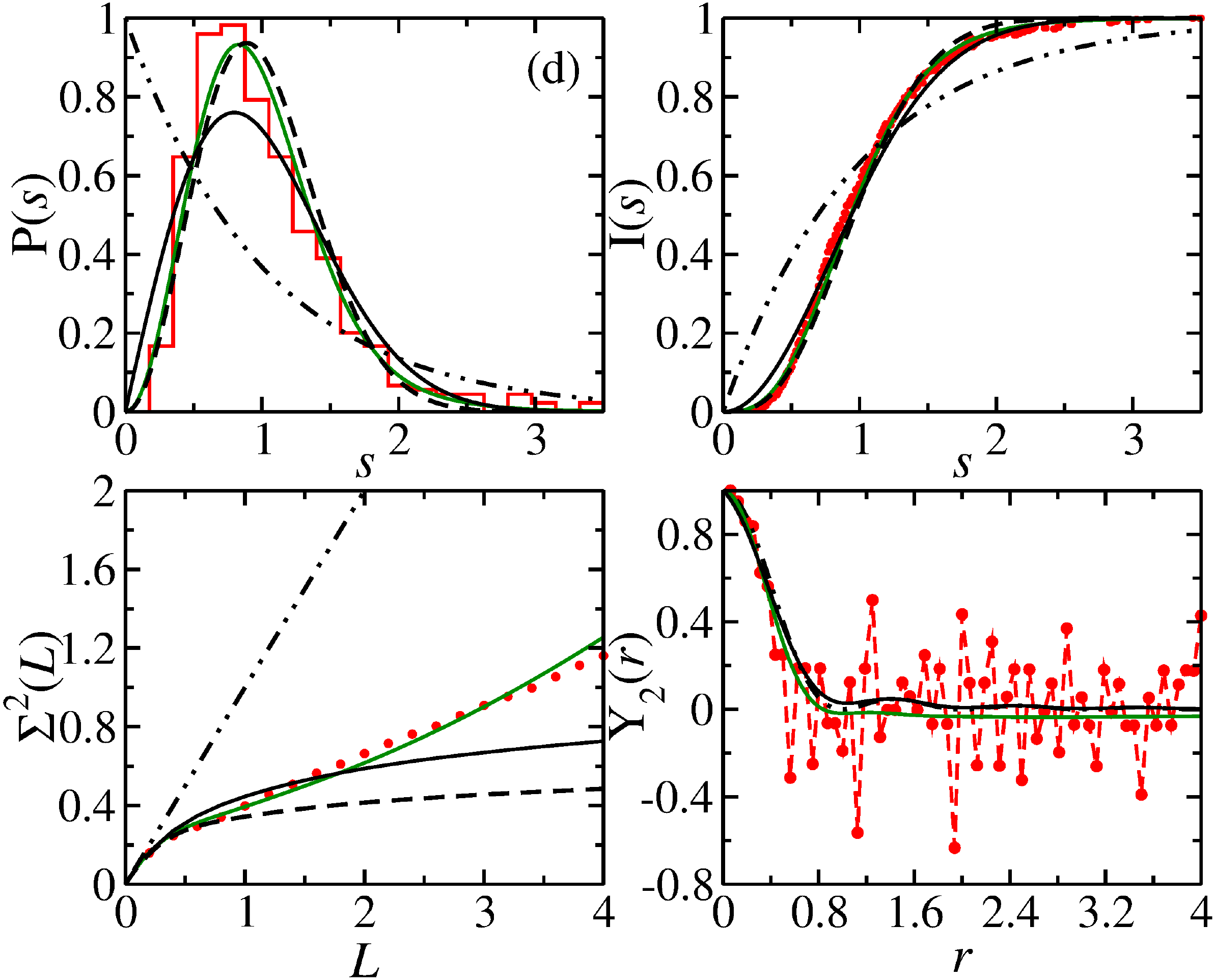}
	\caption{Spectral properties of the cavity {\bf CB1} with $B=200$~mT (red histograms and dots). They are compared to the  curves deduced from Eqs.~(\ref{Y2Anal}),~(\ref{Sigma2Anal}) and~(\ref{PS}) (green curves) and results for Poisson (black dashed-dot-dot lines), GOE (solid black line) and GUE (dashed black lines) statistics. Shown are the results for (a) $n=231$, $f_i\in [10,13]$~GHz, (b) $n=294$, $f_i\in [13,16]$~GHz,  (c) $n=232$, $f_i\in [16,18]$~GHz and (d) $n=256$, $f_i\in [18,20]$~GHz, with $\lambda =0.325,0.45,0.55,0.625$, respectively.}
\label{SpectralPart}
\end{figure}

\begin{figure}[ht!]
        \includegraphics[width=0.8\linewidth]{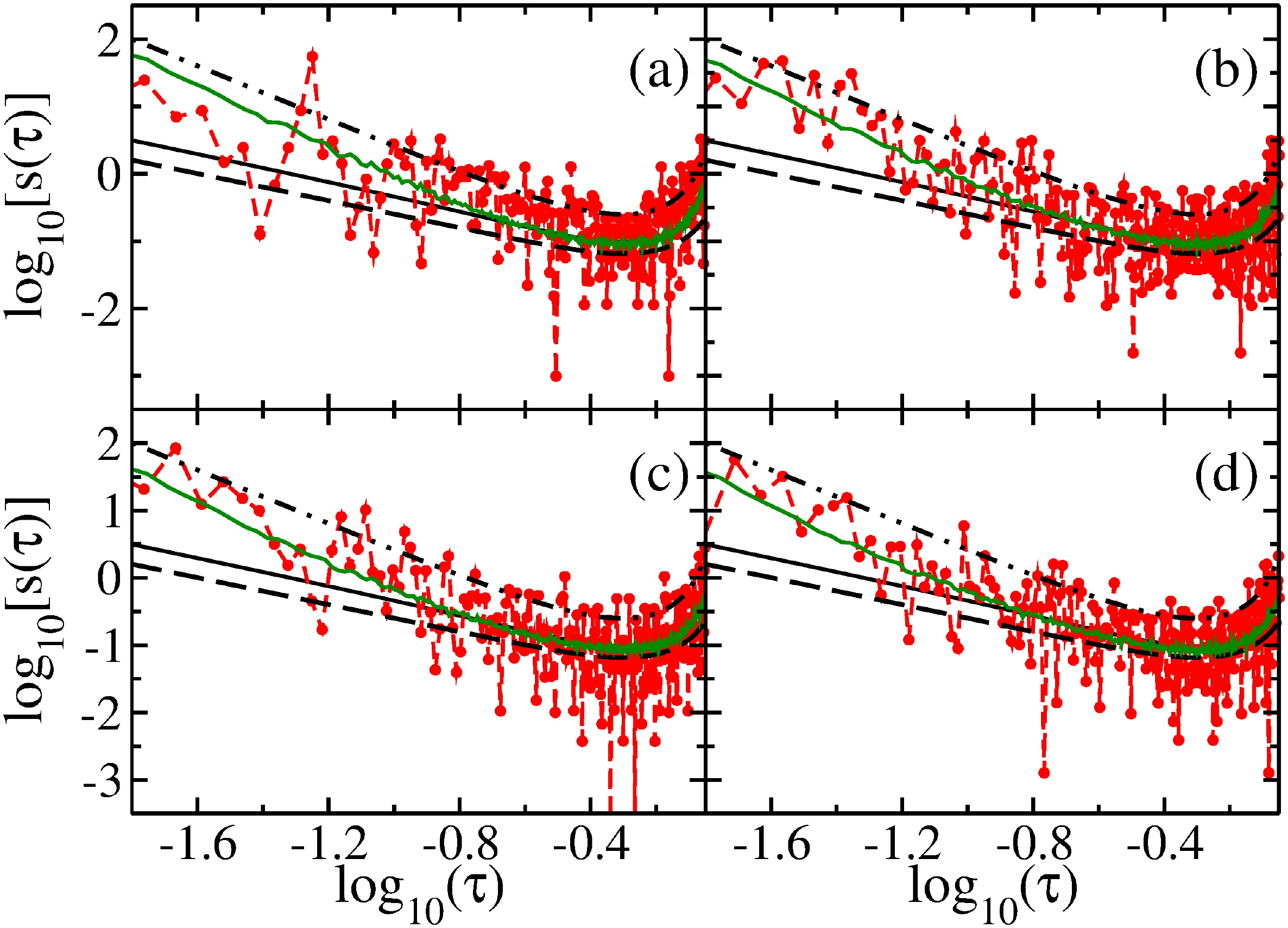}
	\caption{Power spectra for the cavity {\bf CB1} with $B=200$~mT (red histograms and dots). They are compared to curves deduced from Monte-Carlo simulations (green curves) and results for Poisson (black dashed-dot-dot lines), GOE (solid black line) and GUE (dashed black lines) statistics for (a) $n=231$, $f_i\in [10,13]$~GHz, (b) $n=294$, $f_i\in [13,16]$~GHz,  (c) $n=232$, $f_i\in [16,18]$~GHz and (d) $n=256$, $f_i\in [18,20]$~GHz, with $\lambda =0.325,0.45,0.55,0.625$, respectively.}
\label{PowerSpectr}
\end{figure}

Actually, using the whole frequency range from 10-20~GHz corresponds to superimposing spectra with different values of $\lambda$. To demonstrate this, we analyzed the spectral properties in frequency intervals of approximately constant $\Delta_{ab}$ (see~\reffig{SketchSC}). They are shown together with the analytical curves in~\reffig{SpectralPart}. The corresponding values of $\lambda$ are given in the figure caption. Deviations are visible in the long-range correlations beyond a certain value of $L$. This may be attributed to the comparatively small number of eigenvalues $n$ given in the figure caption. The associated ratio distributions are in all frequency ranges similar to the results shown in~\reffig{Ratiod} (b) and (d), implying that they are not sensitive to the changes in the value of $\lambda$, i.e., to the size of chaoticity and \Ti-invariance violation. Furthermore, we compared experimental results for the form factor to the analytical prediction deduced from~\refeq{B2Anal}. In this case we had to cope with the problem that we have sequences of only few hundreds of eigenvalues, however, for the Fourier transform long sequences are preferable. Furthermore, we have only one sequence for each value of $\lambda$, whereas, e.g. in the experiments~\cite{Bialous2016,Che2021} an ensemble of up to a few hundreds of spectra of comparable lengths. So we observe a qualitative agreement with the analytical results confirming the values of $\lambda$, however, these data cannot be used to determine $\lambda$. In~\reffig{PowerSpectr} we compare the experimentally obtained power spectra to Monte-Carlo simulations with the RP model~\refeq{RPH} and to the curves for the GOE and GUE which were obtained based on the analytical expressions in terms of the form factor derived in Ref.~\cite{Faleiro2004}. The smallest value of $\tau$ is $\frac{1}{n}$, with $n$ denoting the number of eigenfrequencies given in the caption. Nevertheless, differences between the power spectra for the different frequency ranges are visible below $\log_{10}(\tau)\simeq -0.5$, and they agree well with the Monte-Carlo simulations.   
\begin{figure}[ht!]
        \includegraphics[width=0.8\linewidth]{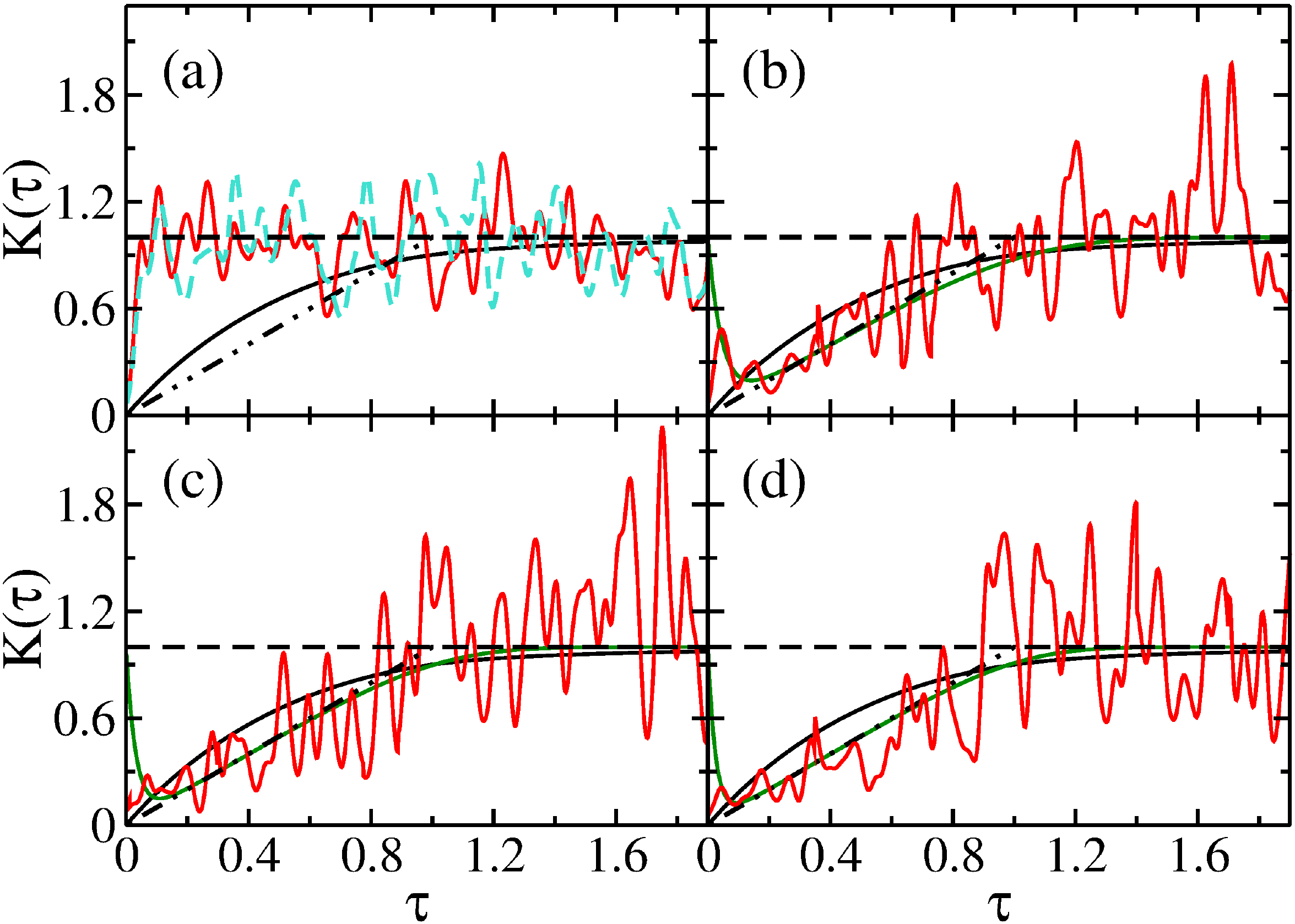}
	\caption{Same as~\reffig{SpectralPart} for the spectral form factor of the cavities (a) {\bf CB1} (red solid line) and {\bf CB2} (turquoise dashed line) for $B = 0$~mT and $\approx 1000$ eigenfrequencies in the frequency range [10,20] GHz and (b)-(d) {\bf CB1} with $B=200$~mT (red solid lines) compared to the analytical results~\refeq{B2Anal} (green curves) and results for Poisson (black dashed-dot-dot lines), GOE (solid black line) and GUE (dashed black lines) statistics. Shown are the results for (b) $n=294$, $f_i\in [13,16]$~GHz,  (c) $n=232$, $f_i\in [16,18]$~GHz and (d) $n=256$, $f_i\in [18,20]$~GHz, with $\lambda =0.45,0.55,0.625$, respectively.} 
\label{SpectralForm}
\end{figure}

\begin{figure}[ht!]
\includegraphics[width=0.8\linewidth]{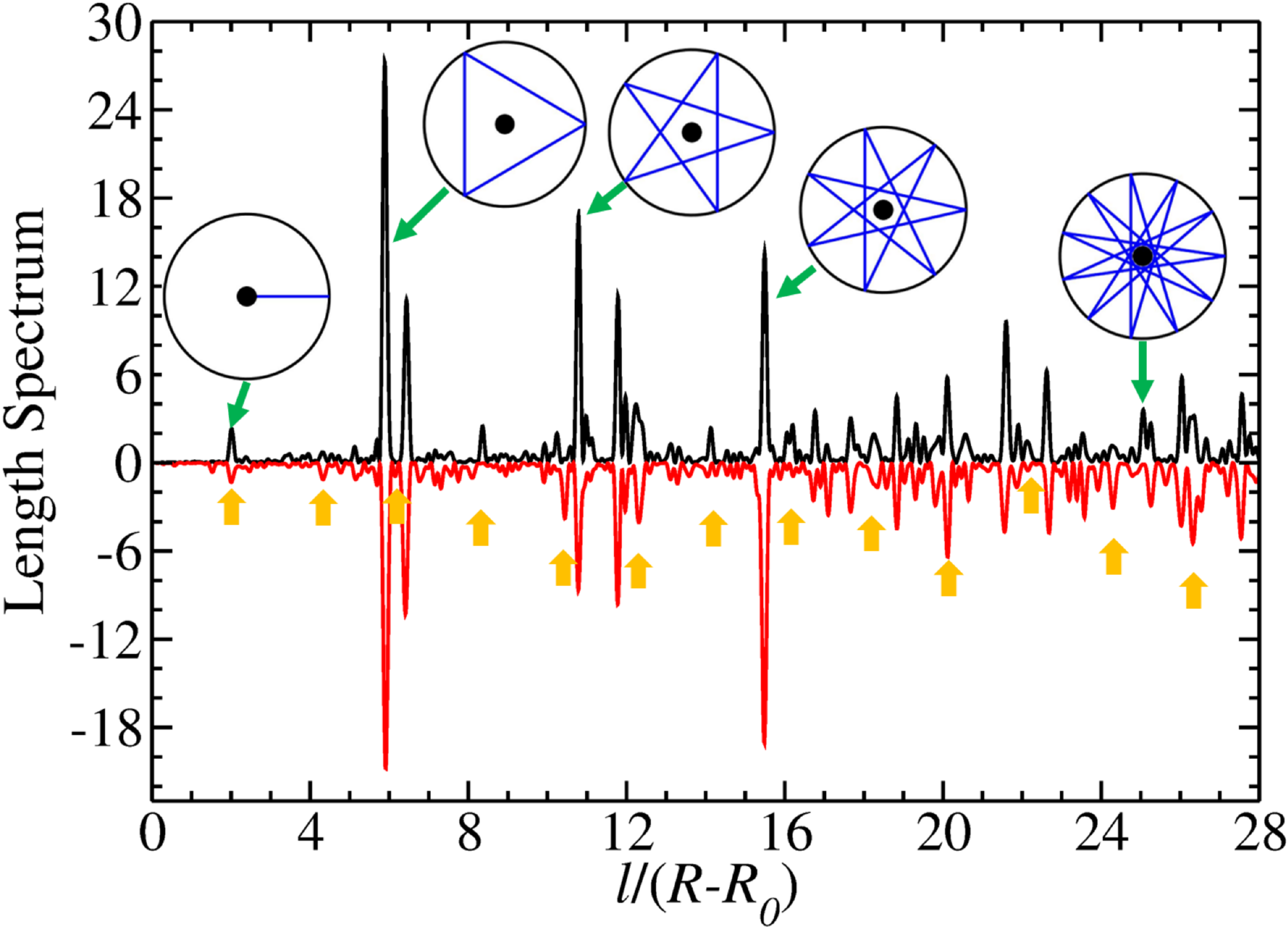}
\includegraphics[width=0.8\linewidth]{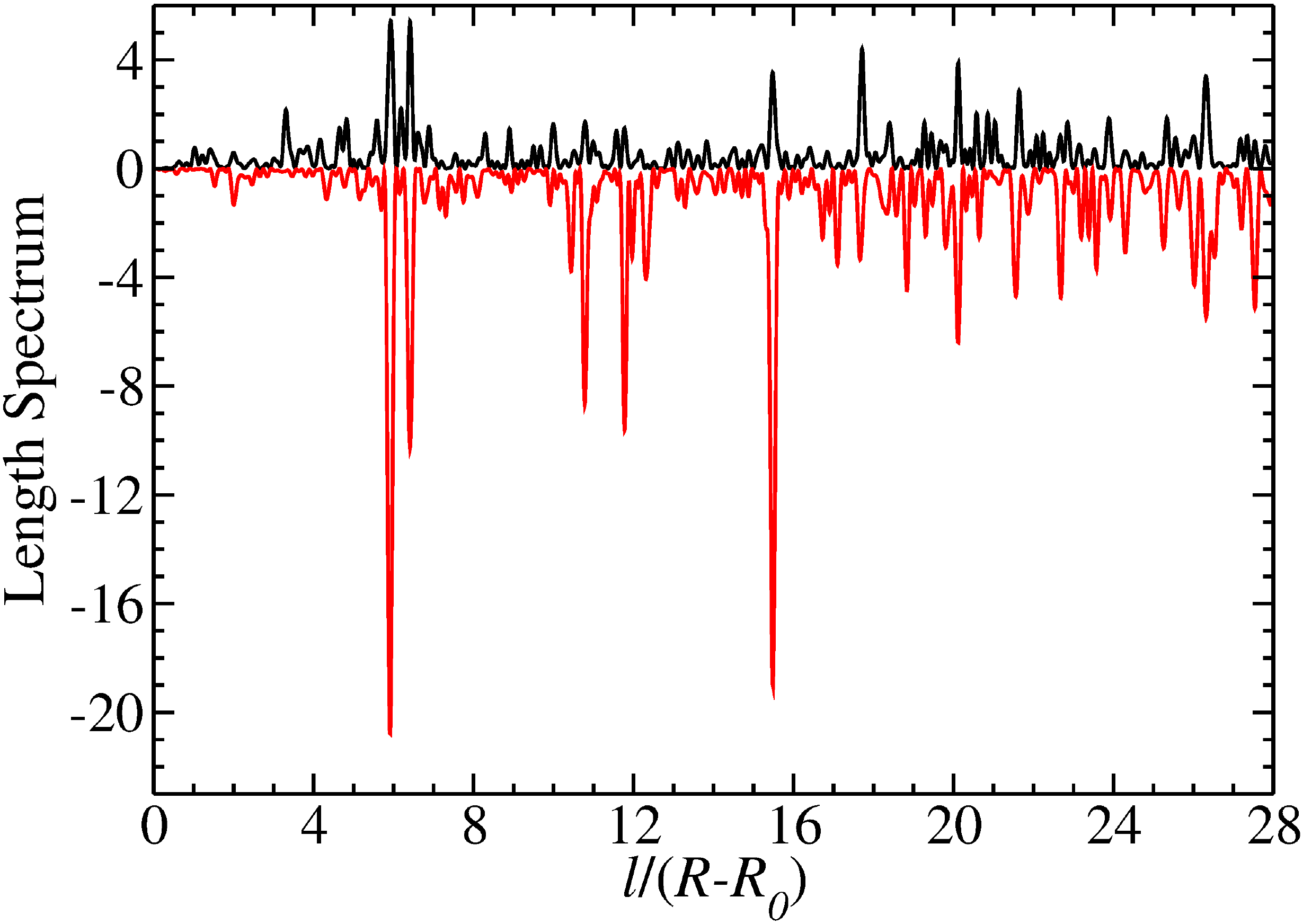}
	\caption{Top: Length spectra of the cavities {\bf CB1} (red solid line) and {\bf CB2} with radius $R=250$~mm containing a ferrite, respectively metallic disk with radius $R_0=30$~mm (black solid line).  Orbits, that hit the ferrite disk at the center of the circular billiard and thus feel the differing boundary conditions, are marked by yellow arrows. Green arrows point at the peaks corresponding to the lengths of the orbits shown in the insets.
	Bottom: Same as top for cavity {\bf CB1} with $B=200$~mT (black solid line) and $B=0$~mT (red solid line).}
\label{FFTs}
\end{figure}

Furthermore, we analyzed length spectra of the three microwave billiard systems. A length spectrum is given by the modulus of the Fourier transform of the fluctuating part of the spectral density from wave number to length and has the property that it exhibits peaks at the lengths of the periodic orbits of the corresponding classical system. The upper part of~\reffig{FFTs} shows the length spectra for the cavities {\bf CB1} with $B=0$~mT and {\bf CB2}. Both length spectra exhibit peaks at the lengths of orbits of the corresponding ring QB. Some peaks are either weakened or suppressed for {\bf CB1}. This is attributed to the differing BCs at the walls of the metallic and ferrite disks, implicating for the latter that in the classical limit there is no specular (hard-wall) reflection at the inner circle. The length spectrum for $B=200$~mT, shown as black curve in the lower part of~\reffig{FFTs}, some peaks are suppressed or disappear, implying that the corresponding periodic orbits do not exist anymore. These are orbits, that hit the disk at the center of the circular billiard, marked by yellow arrows. Furthermore, we show some periodic orbits. Green arrows point at the corresponding peaks. In~\reffig{WFs} of the appendix we show examples of the electric and magnetic field distributions to illustrate the effect of tye magnetized ferrite, which  above its cutoff frequency $f_F^{cut}\approx 4.5$~GHz acts like a random potential~\cite{Zhang2023a}.

\section{Conclusions\label{Concl}}
We propose an experimental setup -- consisting of a flat microwave cavity with the shapes of an integrable billiard, containing ferrite pieces, that are positioned and shaped such that the integrability is not destroyed as long a they are not magnetized -- for the study of the properties of typical quantum systems, whose classical counterpart experiences a transition from integrable with preserved \Ti-invariance to chaotic with partially violated \Ti variance. In~\refsec{RT} we demonstrate in room-temperature experiments with a flat circle-sector shaped cavity, that the fluctuation properties of the $S$ matrix associated with the resonance spectrum of such cavities are well described by the Heidelberg approach~\refeq{Mahaux} with the Hamiltonian replaced by the RP Hamiltonian~\refeq{RPH}. Furthermore, in~\refsec{HeT} we show that the spectral properties of the eigenfrequencies of a circular flat cavity identified in superconducting experiments agree with those of the eigenvalues of the RP Hamiltonian~\refeq{RPH}. We confirmed this by comparing them to and thereby verifying analytical results derived in~\cite{Lenz1992,Kunz1998,Frahm1998}. These experiments were performed with a cavity whose bottom plate and lid are constructed from niobium, a superconductor of type II~\cite{Shubnikov1937}, thereby achieving high-quality factors. This is a crucial prerequisite to render possible the determination of a complete sequence of $\approx 1000$ eigenfrequencies. Thereby, we were able to analyze the spectral properties in various frequency ranges and thus, to observe the gradual transition from Poisson to GUE. Unfortunately, we are not able to measure wave functions with our setup, which relies on Slater's theorem employing a perturbation body made from magnetic rubber~\cite{Bogomolny2006} that is moved along the billiard surface with a guiding magnet, which would interfere in the vicinity of the ferrite with the strong magnetic field magnetizing it. However, there the wave functions show clear distortion from those of the integrable billiard as illustrated in~\reffig{WFs}. A task for the future is to implement another method which doesn't use guideing magnets.     

\section{Acknowledgement} This work was supported by the NSF of China under Grant Nos. 11775100, 12247101 and 11961131009. WZ acknowledges financial support from the China Scholarship Council (No. CSC-202106180044). BD and WZ acknowledge financial support from the Institute for Basic Science in Korea through the project IBS-R024-D1. We thank Sheng Xue Zhang who helped with the design of the niobium parts. XD thanks Junjie Lu, who taught him how to do the experiments. 

XZ and WZ contributed equally to the work.
\bibliography{References}

\begin{appendix}
\renewcommand{\theequation}{A\arabic{equation}}
\renewcommand{\thefigure}{A\arabic{figure}}
\section{Examples of electric and magnetic field distributions}
\begin{figure}[htbp]
        \includegraphics[width=0.8\linewidth]{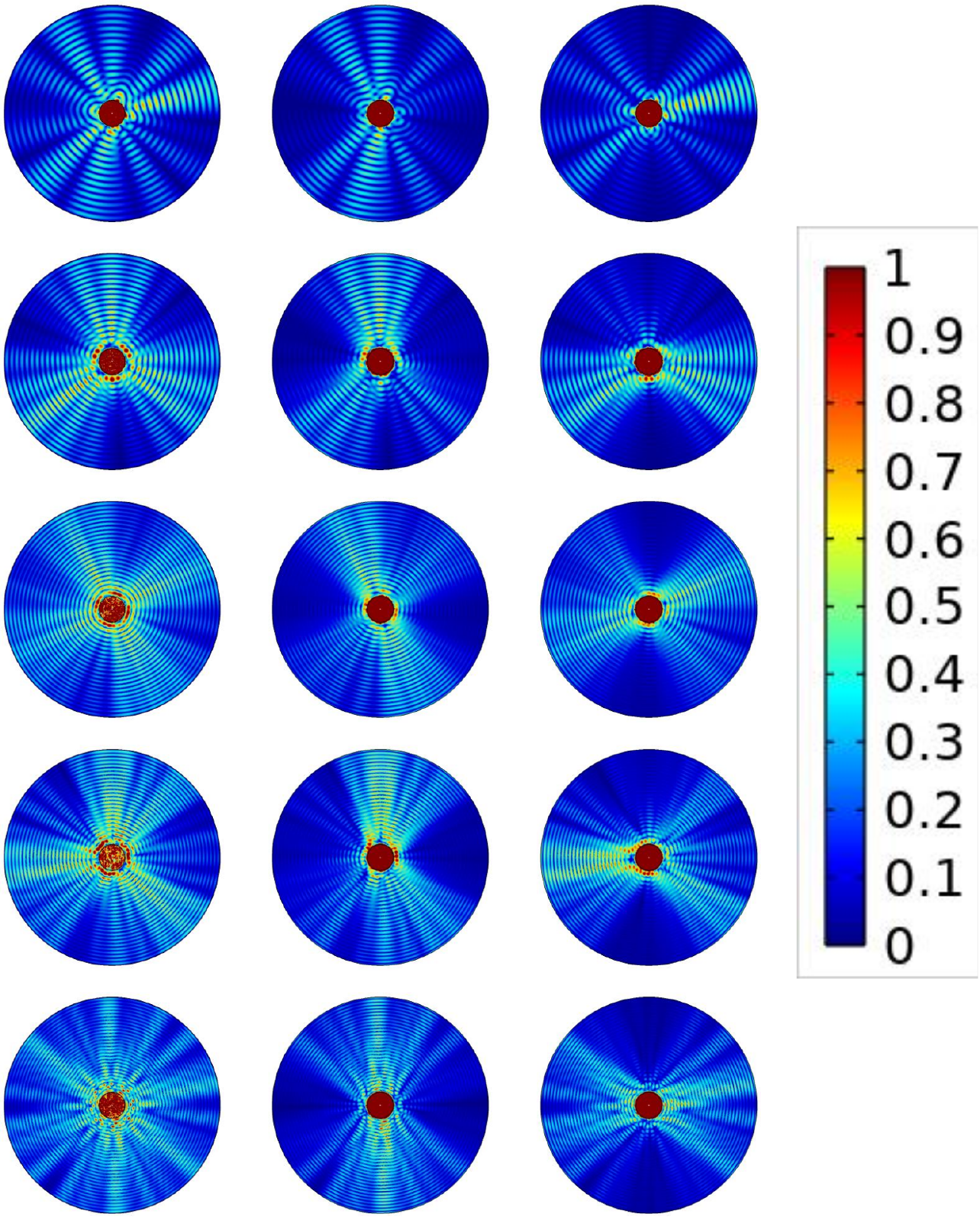}
        \caption{
                Intensity distributions of the electric field component of the electromagnetic waves along the cavity axis, that is in $z$ direction, $\vert E_z\vert$, (first column) and the magnetic field components in $x$ and $y$ direction (second and third column), respectively, for, from top to bottom, $f=10.0012, 12.9405, 16.0522, 17.9931, 19.9404$~GHz.}
        \label{WFs}
\end{figure}

	In~\reffig{WFs} we show examples for the intensity distributions of the electric-field component $E_z$, which for $B=0$~mT corresponds to the wave functions of the ring QB below $f^{cut}=30$~GHz, and for the magnetic-field components $H_x$ and $H_y$ in the cavity with magnetized ferrite. All other field components vanish below $f^{cut}$. The cutoff frequency of the ferrite disk beyond which the electric field distribution becomes three dimensional, equals $f_F^{cut}\approx 4.5$~GHz. We demonstrated in~\cite{Zhang2023a}, that then the wave dynamics of the disk becomes chaotic and \Ti-invariance is completely violated. Thus, above $f=4.5$~GHz it acts like a random potential. This is illustrated in this figure. The distributions were computed with COMSOL Multiphysics. The patterns exhibit clear distortions with respect to those of the corresponding ring-shaped QB. Indeed, as demonstrated for the corresponding spectral properties, the cavity {\bf CB1} exhibits above 10~GHz clear deviations from Poisson statistics and is well described by the RP model for the transition from Poisson to GOE.
\end{appendix}

\end{document}